\documentclass[12pt,linesnumbered,ruled,vlined]{article}
\usepackage{mathptmx}
\usepackage[utf8x]{inputenc}
\usepackage[authoryear]{natbib}
\usepackage{color}
\usepackage{array}
\usepackage{booktabs}
\usepackage{mathtools}
\usepackage{amsmath}
\usepackage{amsthm}
\usepackage{amssymb}
\usepackage{graphicx}
\usepackage{appendix}
\usepackage[unicode=true,pdfusetitle,
 bookmarks=true,bookmarksnumbered=false,bookmarksopen=false,
 breaklinks=false,pdfborder={0 0 1},backref=false,colorlinks=true]
 {hyperref}
\hypersetup{
 pdfborderstyle={},pdfborderstyle={},pdfborderstyle={},urlcolor=blue,linkcolor=blue,citecolor=blue}

\makeatletter

\@ifundefined{date}{}{\date{}}

\usepackage{amsfonts}
\usepackage{dsfont}

\usepackage{geometry}
\usepackage{fancyhdr}
\allowdisplaybreaks[4]
\DeclareMathAlphabet{\mathcal}{OMS}{cmsy}{m}{n}

\usepackage{algorithm}
\usepackage{algorithmic}
\newfloat{listing}{tb}{lst}{}
\floatname{listing}{Listing}
\usepackage{longtable}

\makeatother

\usepackage{authblk}

\begin{document}


\title{Large Language Models at Work in China's Labor Market\small{\thanks{This work is published in \textit{China Economic Review}, Volume 92 (2025), 102413, available at \url{https://doi.org/10.1016/j.chieco.2025.102413}. We are grateful to Jin Li and Xiaodong Zhu for their insightful comments and to the anonymous reviewers for their constructive feedback that helped improve the paper. Qin Chen and Jinfeng Ge acknowledge financial support from the Key Program of the National Natural Science Foundation of China (Grant No. 71933001). Xingcheng Xu and Huaqing Xie are supported by the Shanghai Artificial Intelligence Laboratory. All authors contributed equally to this work. The views expressed in this paper are those of the authors and should not be interpreted as reflecting the views of the National Natural Science Foundation of China or other affiliated institutions. E-mail addresses: qinchen1986@hotmail.com (Q. Chen), jfge@shmtu.edu.cn (J. Ge), xiehuaqing@pjlab.org.cn (H. Xie), xingcheng.xu18@gmail.com (X. Xu), yanqingyang@fudan.edu.cn (Y. Yang).}}}



\author[1]{Qin Chen}
\author[2]{Jinfeng Ge\thanks{Corresponding authors.}}
\author[3]{Huaqing Xie}
\author[3]{\\ Xingcheng Xu$^{\dagger}$}
\author[4]{Yanqing Yang$^{\dagger}$}
\affil[1]{MetroData, China}
\affil[2]{School of Economics \& Management, Shanghai Maritime University}
\affil[3]{Shanghai Artificial Intelligence Laboratory}
\affil[4]{Fudan University}


\maketitle

\begin{abstract}

This paper explores the potential impacts of large language models (LLMs) on the Chinese labor market. We analyze occupational exposure to LLM capabilities by incorporating human expertise and LLM classifications, following the methodology of \cite{eloundou2023gpts}. The results indicate a positive correlation between occupational exposure and both wage levels and experience premiums at the occupation level. This suggests that higher-paying and experience-intensive jobs may face greater exposure risks from LLM-powered software. We then aggregate occupational exposure at the industry level to obtain industrial exposure scores. Both occupational and industrial exposure scores align with expert assessments. Our empirical analysis also demonstrates a distinct impact of LLMs, which deviates from the routinization hypothesis. We present a stylized theoretical framework to better understand this deviation from previous digital technologies. By incorporating entropy-based information theory into the task-based framework, we propose an AI learning theory that reveals a different pattern of LLM impacts compared to the routinization hypothesis.
\end{abstract}

\section{Introduction\label{sec:Introduction}}

Recent remarkable progress in the field of generative AI and large language models~\citep{bubeck2023sparks,zhao2023survey} has provoked many pressing questions about the effects of these powerful technologies on the economy. One of the most significant questions surrounding advances in generative AI and LLMs is the impact these technologies will have on the dynamics of the labor market due to the influence of LLMs on labor inputs. A branch of research emphasizing the disruptive labor market impacts of LLMs is emerging rapidly, however, it predominantly focuses attention on the labor market in developed economy, in particularly the U.S.~\citep{eloundou2023gpts,peng2023impact,noy2023experimental,brynjolfsson2023generative,felten2023occupational}. Nevertheless, countries differ in their labor market structures such as occupation and industry composition. Even for the same occupation, the detailed task composition or work content may exhibit great discrepancies across countries. Therefore, this paper analyzes the potential impacts of LLMs on China's labor market.
To construct our primary exposure index, we use a recently developed methodology to systematically assess which occupations are most exposed to LLMs in China. Specifically, we employ three large language models -- GPT4~\citep{OpenAI2023GPT4TR}, InternLM~\citep{2023internlm}, and GLM~\citep{zeng2022glm,du2022glm} -- as classifiers to determine the occupational exposure based on the detailed description for each occupation contained in the general code of occupational classification of the People's Republic of China. We also employ expert annotators to explore the impacts of LLMs, to make comparisons, and to shed more light on this issue. 
Our key empirical findings are twofold. First, we empirically analyze occupational exposure to LLMs, highlighting the implications of heterogeneous occupational exposure. Second, we leverage these findings to assess the labor market demand exposure to LLMs across industries and demographic groups in China.

The analysis reveals significant heterogeneity in occupational exposure, with more educated, higher paid, white-collar occupations being the most exposed to LLMs. These results align with recent studies on the US labor market ~\citep{eloundou2023gpts,eisfeldt2023generative}, which also found that technological advancements disproportionately impact workers at the higher end of the wage spectrum. Beyond the positive correlation between wage, education, and occupational exposure, our findings indicate a positive correlation between experience premiums and exposure to LLMs, suggesting potential diminishing returns to ``learning by doing" in the future.

To deepen these insights, we employed advanced language models (e.g., GPT-4, GLM, InternLM) to evaluate the quintet of occupations defined by \cite{acemoglu2011skills}: Non-routine Cognitive (Analytical), Non-routine Cognitive (Interpersonal), Routine Cognitive, Routine Manual, and Non-routine Manual Physical. For each occupation, we calculated an aggregate exposure score by averaging sub-occupation evaluations. The results indicate that non-routine cognitive skills, particularly analytical ones, are most significantly affected by LLMs. Routine manual skills are also substantially influenced, whereas interpersonal and non-routine manual physical skills remain relatively unaffected.

This predominant impact on non-routine cognitive tasks deviates from the ``routinization hypothesis" ~\citep{autor2003skill}, which posited that ICT primarily automates routine, codifiable tasks performed by middle-wage workers. To explore the different occupational exposure patterns between traditional AI technologies and LLMs, we developed a theoretical model to shed light on the underlying mechanisms and policy implications.

Our analysis further indicates that the impacts of LLMs on China's labor market are pervasive and diverse. Industry-level analyzes reveal that education and healthcare have a higher level of exposure, while manufacturing, agriculture, mining, and construction show lower exposure. In contrast to other developed countries, the uneven age distribution across industries in China amplifies the demographic exposure to LLMs, disproportionately affecting younger workers.

Furthermore, LLMs significantly impact labor demand. Using an online job posting dataset, we constructed an occupational vacancy index. A positive correlation between vacancy shares and occupational exposure scores suggests that labor demand structures may exacerbate the disruptive impacts of LLMs on China's labor market. The positive correlation between the growth rate of vacancy shares and exposure scores further indicates a potential reversal in labor demand trends. Contrary to expectations, China's economic and labor market structure intensifies rather than mitigates the disruptive effects of LLMs.

Motivated by the unique occupational exposure of LLMs compared to earlier technologies, we introduce a novel theoretical model to examine why different technologies produce different occupational exposure structures. This model incorporates entropy-based information theory into a stylized task-based framework, which is particularly effective for characterizing comparative advantage. By defining task complexity using entropy and modeling the relative productivity of skills across occupations, we can systematically assess the efficiency of AI technologies, including LLMs, across various tasks.

The model also integrates KL-divergence to represent the relative efficiency of AI models across occupations, distinguishing the automation logic of traditional AI from that of deep learning-driven LLMs. This represents a significant advancement in AI technology. Finally, we discuss the theoretical and policy implications of LLMs inferred from our model, providing critical insights for navigating the future of labor markets.

The paper is structured as follows: Section 2 reviews the related literature, Section 3 discusses methods
and data collection, Section 4 presents the main  results, Section 5 introduces a theoretical model to further discuss the impacts of LLMs, and Section 6 offers concluding remarks.

\section{Literature Review\label{sec:Related}}

Artificial Intelligence, like previous technologies, is poised to impact the economy in various ways, potentially fostering economic growth and reshaping the labor market structure \citep{furman2019ai,goldfarb2023could}. A substantial and expanding body of literature delves into the labor market consequences of artificial intelligence and automation technologies broadly defined. The skill-biased technological change framework \citep{katz1992changes,acemoglu2002technical}, along with the task-based framework of automation \citep{autor2003skill,acemoglu2011skills,acemoglu2018race} are often regarded as the foundational frameworks for comprehending technology's impact on the labor market. This line of research has introduced the concept of routine-biased technological change, indicating that workers engaged in routine tasks face a heightened risk of displacement due to technological advancements. Numerous studies have demonstrated that automation technologies have contributed to both income inequality and job polarization, driven by declines in relative wages and employment for workers specializing in routine tasks \citep{autor2006polarization,van2011wage,acemoglu2022tasks}. The influence of AI on work is anticipated to be multi-faceted. Recent studies have made distinctions between task-displacement and task-reinstatement effects of technology, whereby new technology introduces novel occupations that bolster labor demand \citep{acemoglu2018race,acemoglu2019automation}.

Historically, prior research has predominantly adopted the task-oriented approach to analyze the labor market impacts of artificial intelligence. Various methods have been employed to evaluate the similarity between AI capabilities and the tasks performed by workers across different occupations.  These methodologies encompass aligning AI capabilities with diverse skills and abilities demanded by distinct occupations \citep{felten2018method,tolan2021measuring}, mapping AI patent descriptions to worker task descriptions \citep{webb2019impact,meindl2021exposure}, employing machine learning classifiers to estimate the potential for automation across all occupations \citep{frey2017future}, devising innovative rubric the suitability of worker activities by machine learning \citep{brynjolfsson2018can}, and leveraging expert forecasts \citep{grace2018will}. 

Nevertheless, this line of research is becoming increasingly challenging due to the evolving and advancing capabilities of AI. Following recent breakthroughs in Generative AI and LLMs, there has been a growing body of studies investigating the specific economic impacts and opportunities presented by LLMs. For instance, \citep{peng2023impact} conducted a study where software engineers were enlisted for a specific coding task, revealing that those with access to GitHub Copilot completed the task twice as quickly. Similarly, \citep{noy2023experimental} conducted an online experiment to explore the displacement effects of Generative AI on professional writing tasks. Additionally, \citep{brynjolfsson2023generative} examined the effects of Generative AI on customer support agents. Pertinent to this paper, \citep{felten2023occupational} explored the heterogeneity in occupational exposure, while \citep{eloundou2023gpts} introduced a novel rubric to assess the impacts of LLMs on labor forces. Concurrently with this line of inquiry, we aim to characterize the potential relevance of LLMs to China's labor market in particular.

Our theoretical model is based on the task-based automation framework \citep{autor2003skill,acemoglu2011skills,acemoglu2018race}. To better capture the unique characteristics of LLMs, we incorporate insights from information theory into this framework.

Information theory, originating from Shannon's seminal work \citep{shannon2001mathematical}, has been widely applied across various disciplines e.g.\citep{cover1999elements,mackay2003information}. The emergence of deep learning has further expanded its relevance by enabling the representation of complex, nonlinear, and hierarchical data structures. Shannon’s maxim of “information first, then computation” inspires the integration of information theory into AI research, exemplified by the information bottleneck framework \citep{tishby2000information,tishby2015deep}. This framework formulates a learning objective grounded in information theory and prescribes algorithms to optimize it. Information theory has significantly advanced the understanding of deep neural networks, from practical applications like the variational information bottleneck \citep{alemi2016deep} to theoretical investigations of generalization bounds using mutual information \citep{xu2017information}. Recent work by \cite{shwartz2024compress} unifies research on self-supervised and semi-supervised learning through an information-theoretic perspective, discussing optimal representations for neural networks. Notably, information theory has been employed to analyze LLM performance through ``neural scaling laws” \citep{kaplan2020scaling,hoffmann2022training,barnett2023scaling}, which describe the relationship between compute, data, and model performance.

In econometrics, statistics, and economics, information theory has also proven valuable. Its core concept, entropy, quantifies uncertainty in random variables and complexity in transmitted information. \cite{sims2003implications} introduced a dynamic model of rational inattention, where agents face entropy-based costs in processing information. This approach, further developed by \cite{mackowiak2023rational}, provides a flexible framework for modeling information constraints, with entropy serving as a tractable benchmark.

By integrating information theory into our task-based framework, we establish deeper insights into task automation and artificial intelligence. This approach allows us to characterize the relative productivity of skills and AI models across tasks using key concepts such as entropy, KL-divergence, and neural scaling laws. These tools enable a rigorous analysis, akin to the Cobb-Douglas production function, capturing the fundamental distinctions between traditional and advanced AI-driven automation.

\section{Methodology of Exposure Scoring and Data Collection\label{sec:Methods}}

\begin{figure}[htb]
\begin{centering}
\includegraphics[width=0.5\linewidth]{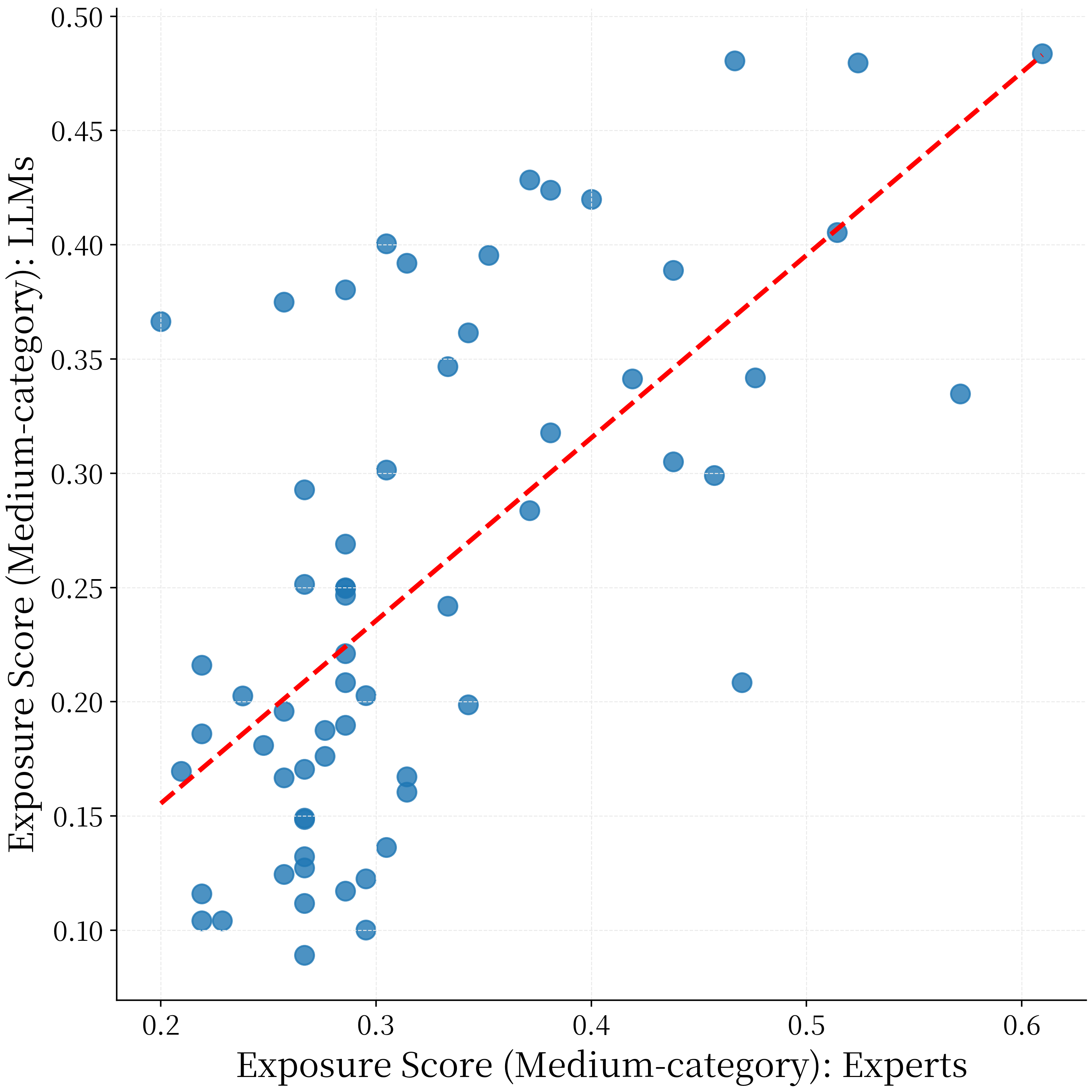}
\par\end{centering}
\caption{Exposure Score of LLMs and Expert: corr.=0.65.}
\label{fig:cor_GLM2+Internlm+GPT_Expert}
\vspace{-1.2em}
\end{figure}

\begin{figure*}[htb]
\begin{centering}
\includegraphics[width=0.95\linewidth]{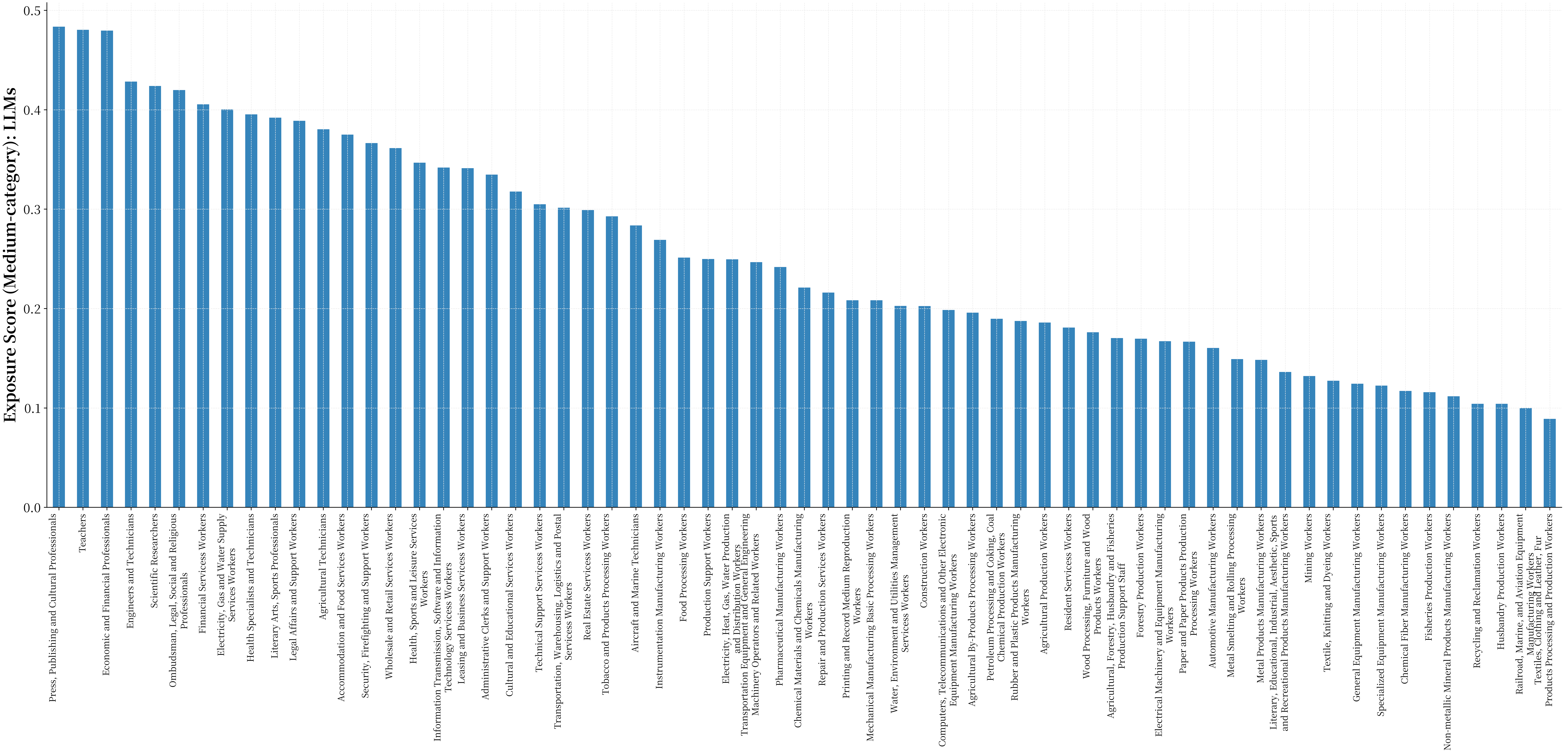}
\par\end{centering}
\caption{Exposure Score on Medium-category Level: LLMs.}
\label{fig:medium_score_GLM2+Internlm+GPT}
\end{figure*}

\begin{figure}[htb]
\begin{centering}
\includegraphics[width=0.65\linewidth]{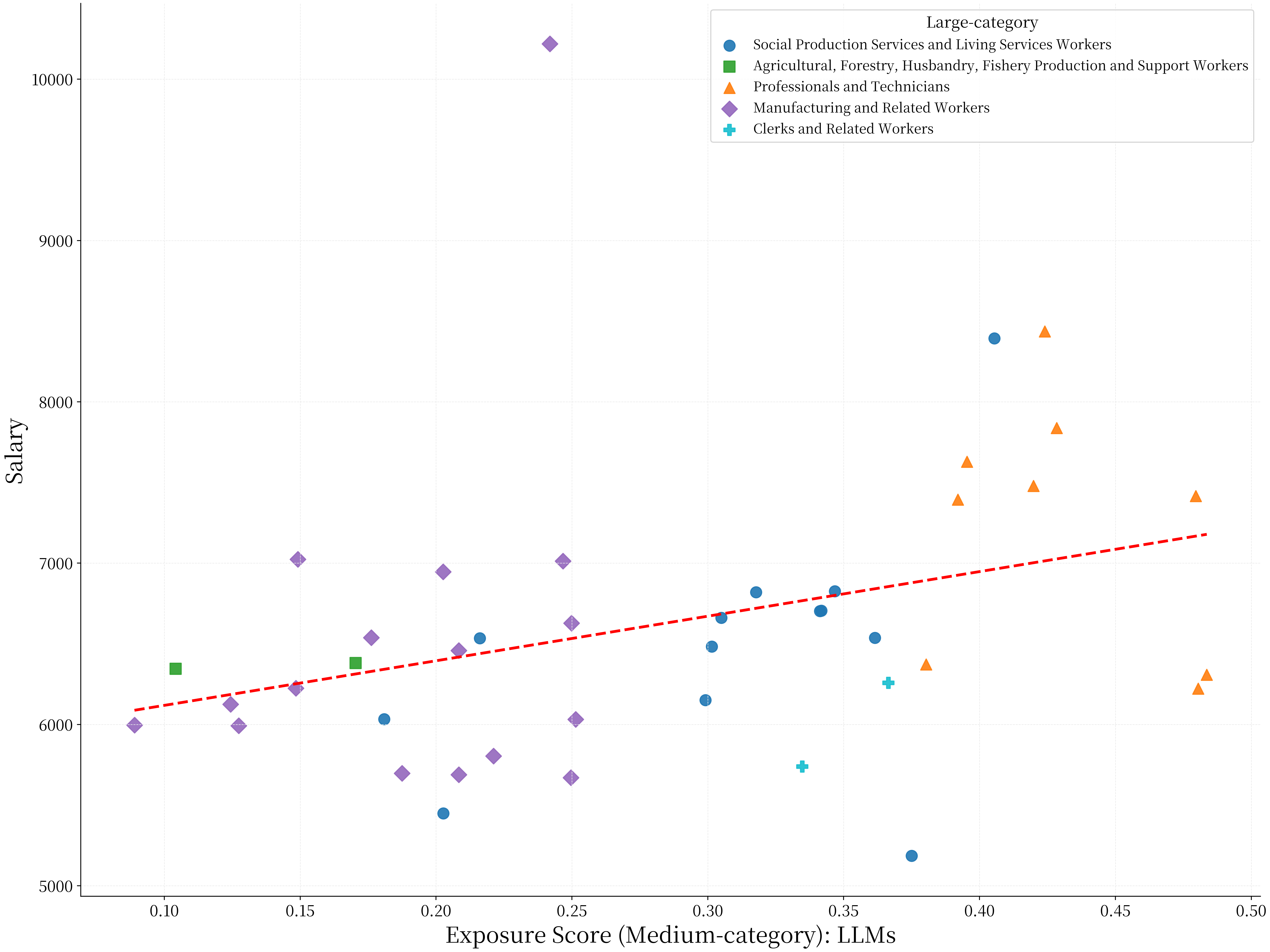}
\par\end{centering}
\caption{Salary and Exposure Score (Medium-category): LLMs.}
\label{fig:salary_GLM2+Internlm+GPT}
\vspace{-1.2em}
\end{figure}
\begin{figure}[htb]
\begin{centering}
\includegraphics[width=0.65\linewidth]{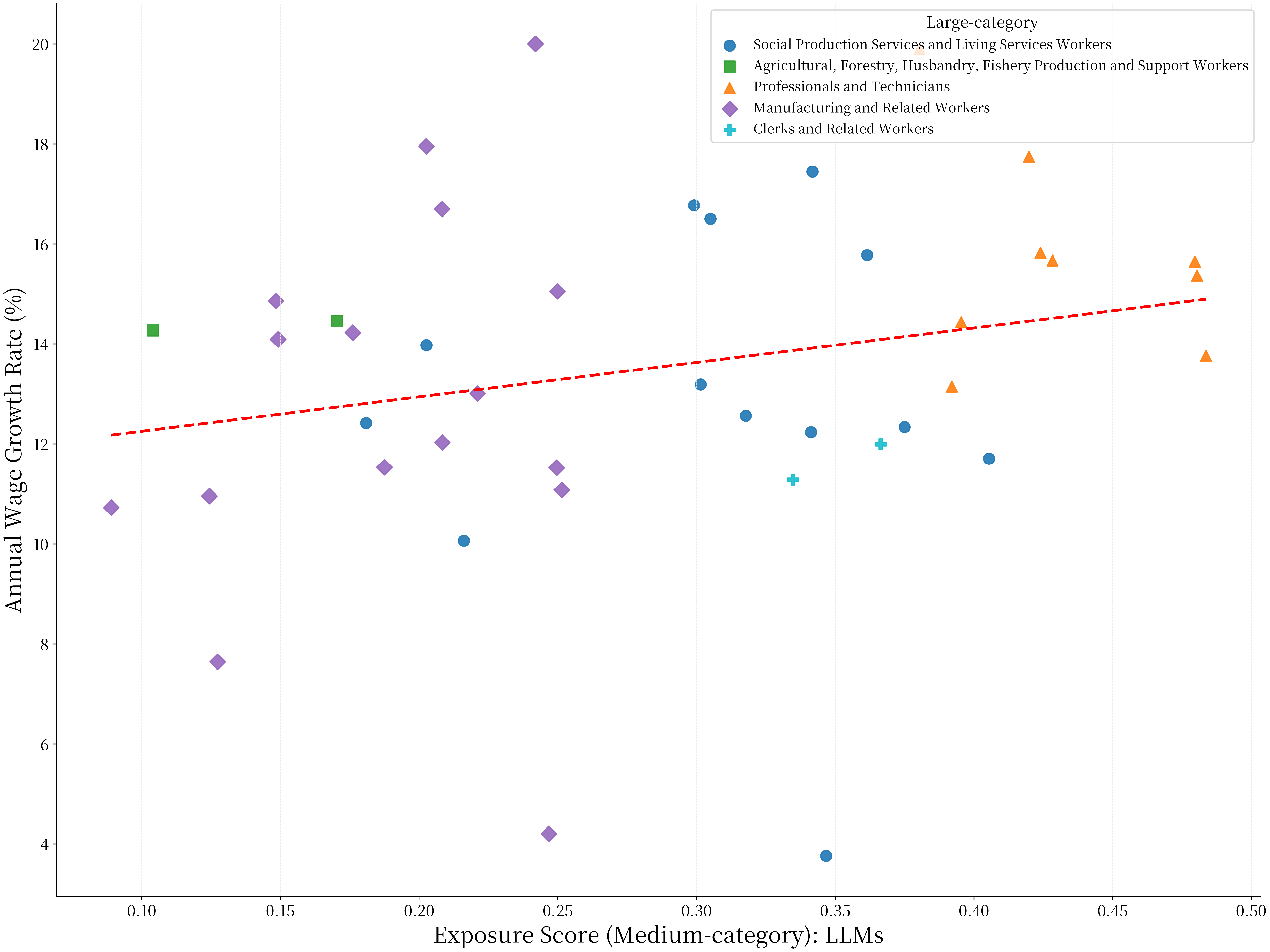}
\par\end{centering}
\caption{Annual Wage Growth Rate and Exposure Score (Medium-category): LLMs.}
\label{fig:wage_growth_GLM2+Internlm+GPT}
\vspace{-1.2em}
\end{figure}
\begin{figure}[htb]
\begin{centering}
\includegraphics[width=0.65\linewidth]{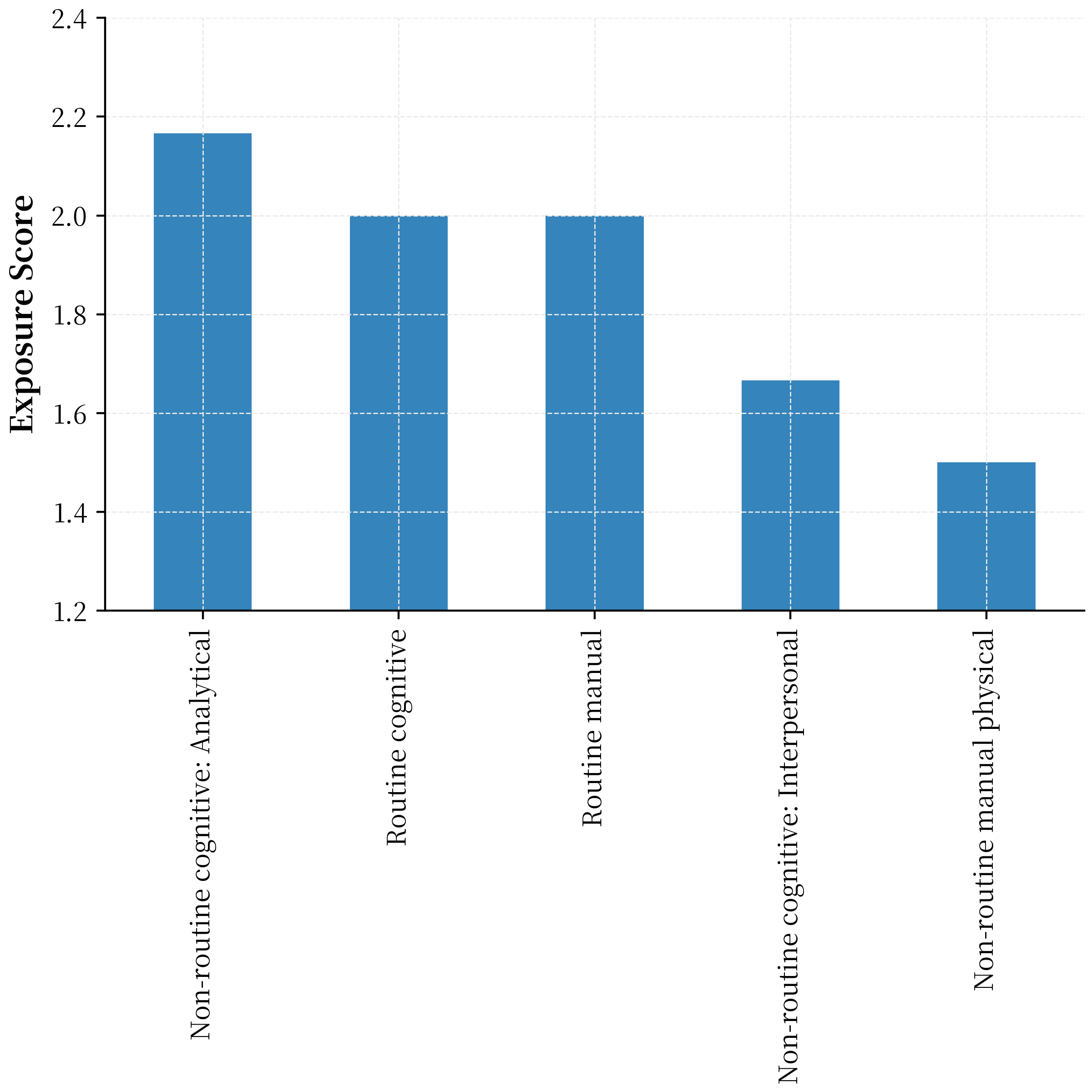}
\par\end{centering}
\caption{Composite Occupation and LLMs exposure.}
\label{fig: skills and llms}
\end{figure}

\begin{figure}[htb]
\begin{centering}
\includegraphics[width=0.75\linewidth]{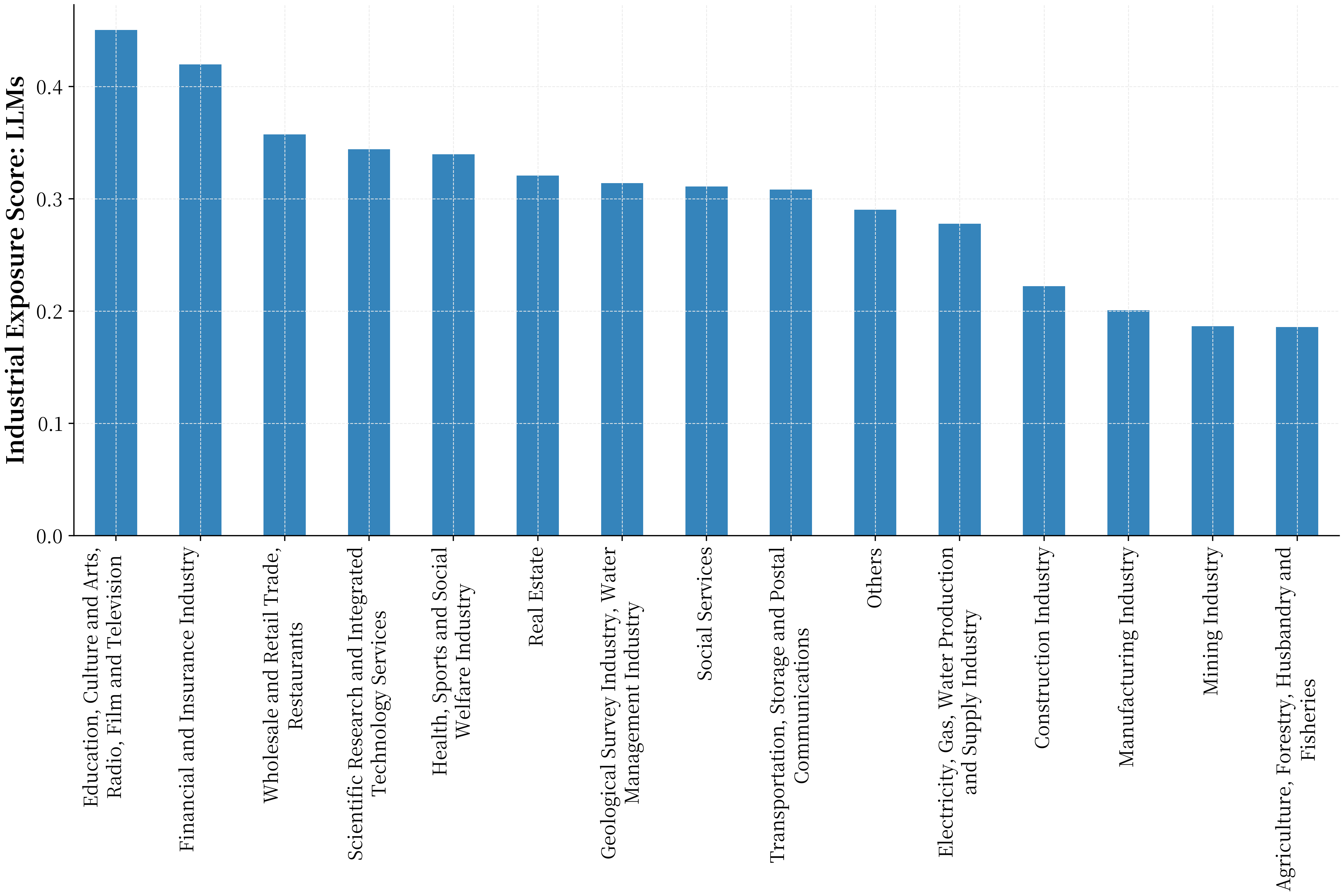}
\par\end{centering}
\caption{Industrial Exposure Score: LLMs.}
\label{fig: ind_score_GLM2+Internlm+GPT}
\end{figure}
\begin{figure}[htb]
\begin{centering}
\includegraphics[width=0.65\linewidth]{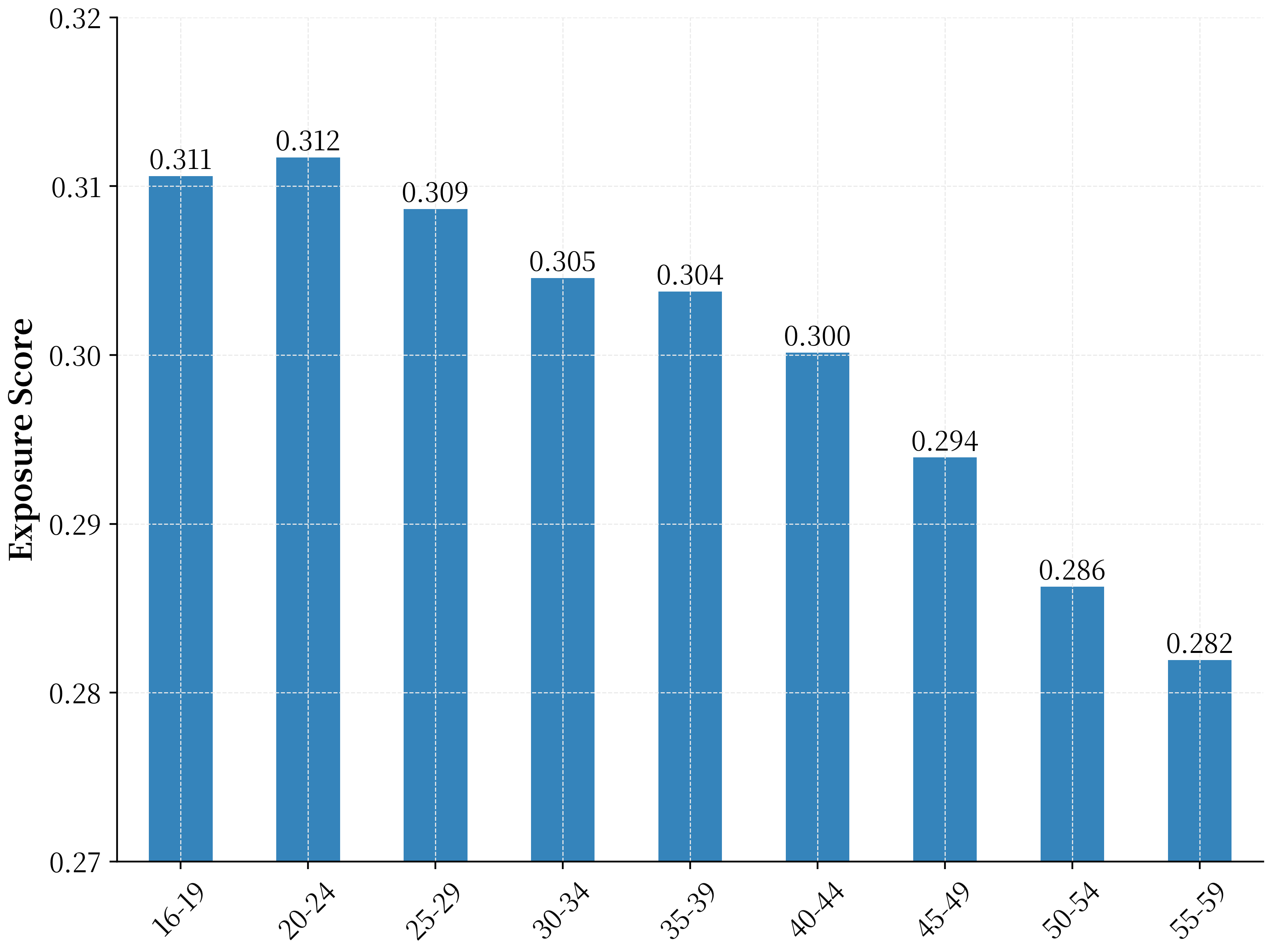}
\par\end{centering}
\caption{Demographic Exposure Score: LLMs.}
\label{fig: demographic exposure}
\end{figure}
\begin{figure}[htb]
\begin{centering}
\includegraphics[width=0.65\linewidth]{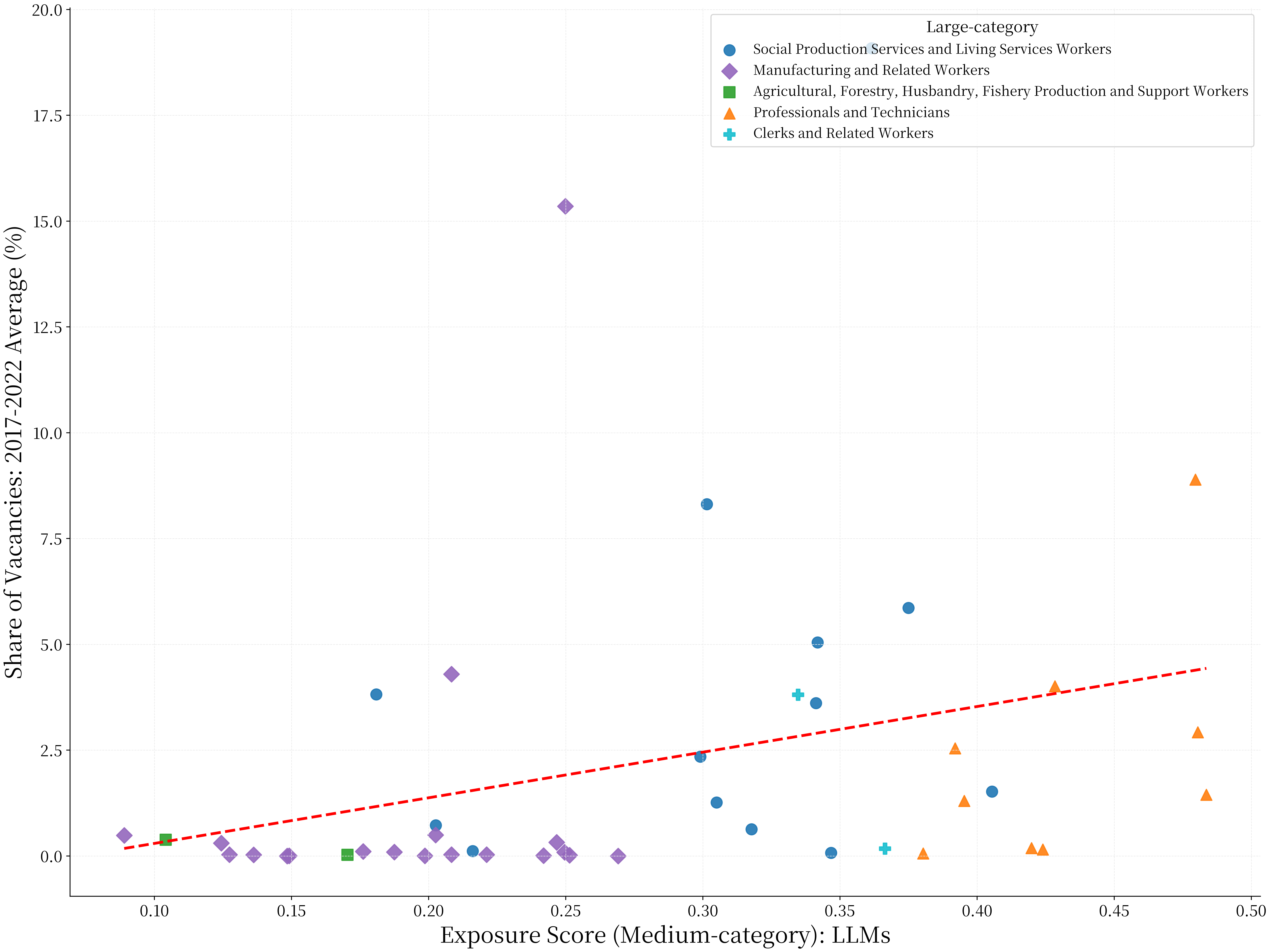}
\par\end{centering}
\caption{Share of Vacancies and Exposure Score(2017-2022): LLMs.}
\label{fig: share of vacancy llms}
\end{figure}
\begin{figure}[htb]
\begin{centering}
\includegraphics[width=0.65\linewidth]{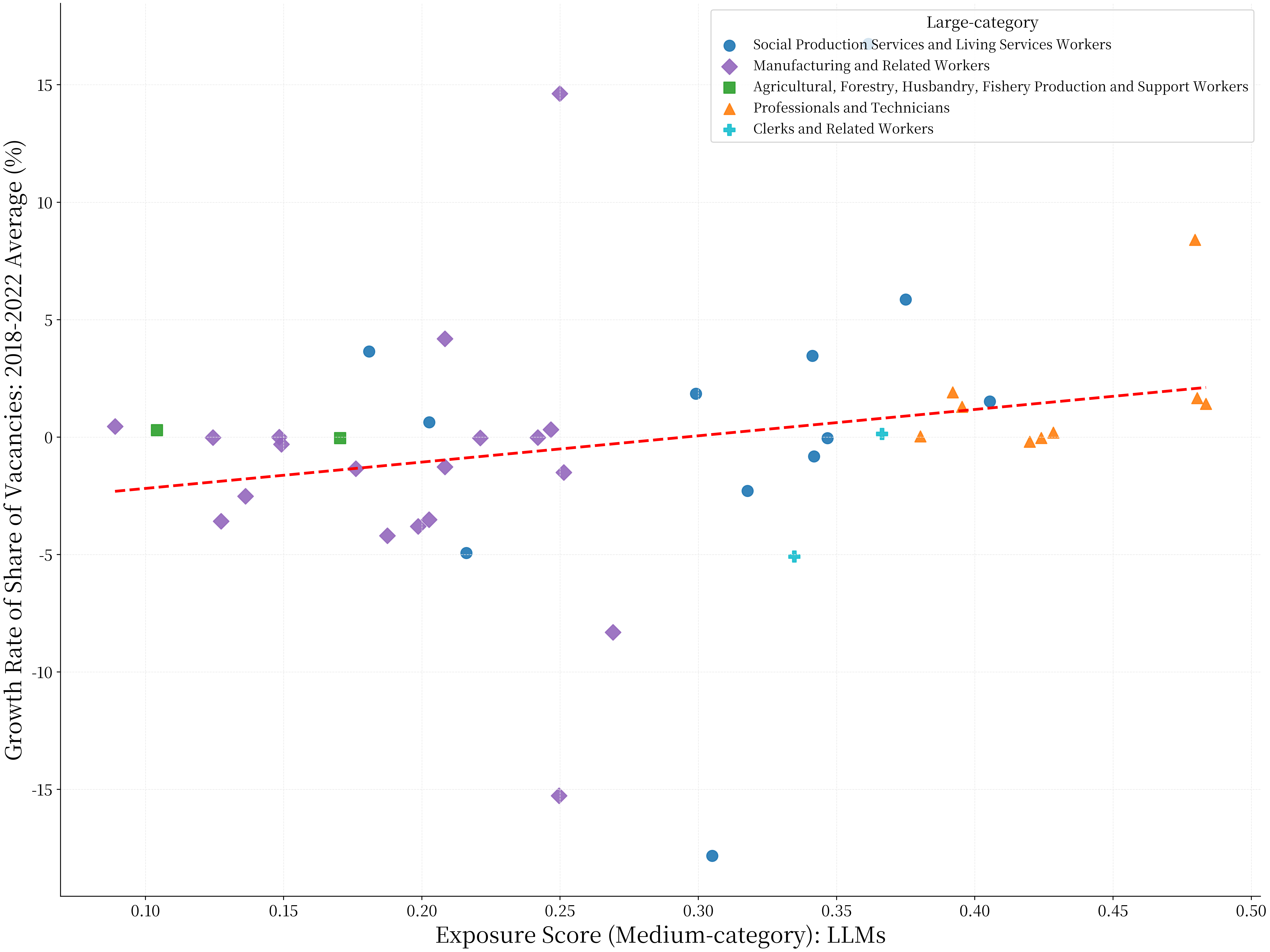}
\par\end{centering}
\caption{Growth Rate of Share of Vacancies and Exposure Score: LLMs.}
\label{fig: percentage change of share of vacancy llms}
\vspace{-1.2em}
\end{figure}

\subsection{Data on Occupations in China}
In order to assess occupational exposure scoring in China and facilitate standardized comparisons, a consistent occupational classification system is essential. The Occupational Classification Dictionary of the People's Republic of China (OCD) version 2022, published by the National Bureau of Statistics of China (NBS) and the Ministry of Human Resources and Social Security of the People's Republic of China (MOHRSS), provides such a classification system that serves as a standardized analytical tool for occupations. In particular, we utilize the general code of Occupational Classification of the People's Republic of China 2022, which encompasses comprehensive information across 8 large categories, 79 medium categories, 449 small categories, and 1636 fine categories. This information, includes the definition of each occupation, the content and format of work activities, as well as specific description of the scope of work activities. We leverage the detailed descriptions of various occupations to facilitate the classification of online job vacancies into distinct occupational categories. 
To provide further clarity, a sample of occupations and their exposure to LLMs, categorized by medium categories, is presented in Appendix A.

\subsection{Data on Wages and Vacancies}

To acquire both vacancy and wage information, we leverage two datasets. Our primary data source is an online job posting dataset collected by the City Data Group. The dataset compiles online job postings spanning from January 2017 to December 2022, originating from major online job market platforms in China including zhaoping.com, 51job, 58.com, Ganji.com, Lagou.com, and Kanzhun. This comprehensive compilation encompasses over 800 million job openings across nearly 400 cities and 5.2 million employers. For each job vacancy entry within this database, we have access to a range of information: including the posting date, position type, occupation titles, the quantity of workers to be recruited, wage ranges (if specified), education requirements (if applicable), work experience prerequisites (if indicated), the name of the employing firm, the work location for the position, and the textual content of job descriptions. Following the classification of each job posting distinct occupations, we derive corresponding statistics encompassing the number of vacancies, the typical educational qualifications required for entry, and the wage structure within each occupation.
The City Data Group frequently reviews the representativeness of the job vacancy postings it scrapes, to ensure the information renders an accurate picture at least the online recruitment labor market. Obviously, the online recruitment labor market does not equal to the whole labor market; to gauge the nuance difference between online recruiting labor market and the overall labor market, the one precondition is a representative aggregate recruitment dataset. Since in China there is no representative recruitment dataset such as JOLTS in United States, therefore, we cannot speak too much about representativeness of online job recruitment compared with the overall recruitment.  Although the potential unknown biased feature of job posting dataset, we believe large proportion of our main empirical results are immune to this potential bias. First, our key empirical task which is eliciting occupational exposure does not use online job posting data at all. We use online posting data to deliver three key information: vacancy numbers, the average wage rate and experience premium for each occupation.  If different sectors are different reliance on online recruitment, indeed the more online recruitment reliant sector’s occupation will overestimate. However, the calculation of the average wage rate and experience premium for each occupation are robust to heterogeneous sectoral reliance on online recruitment.

Our second data source is the China Labor-force Dynamic Survey (CLDS) 2016, conducted by the Social Science Survey Center of Sun Yat-sen University. The CLDS surveys the working-age population to explore aspects such as education, employment, labor rights, occupational mobility, health, and well-being. The survey encompasses comprehensive industry and occupation information for each employment entry. Consequently, we can harness this dataset to construct an occupational intensity index for each industry. This index facilitates the acquisition of an exposure score at the industry level derived from occupational exposure scores. Further details regarding the occupational intensity index across 15 industry categories are presented in Appendix C.

\subsection{Methodology of Exposure Scoring}

To evaluate the likelihood of an occupation in China undergoing a disruptive shock due to the widespread availability and utilization of LLMs, we continue to adopt a task-based approach. We gauge the exposure of each occupation to LLMs based on the comprehensive occupation descriptions in Chinese, as documented the Occupational Classification Dictionary of the People's Republic of China 2022. Building upon the exposure scoring methods proposed by \citep{eloundou2023gpts} which conceive an occupation as a collection of tasks, assess whether a given occupation can be executed more efficiently using ChatGPT or analogous LLMs. Our methodology employs three prominent large language models to determine exposure of various occupations. Specifically, we utilize Open AI's GPT-4 model \citep{OpenAI2023GPT4TR}, the InternLM model developed by Shanghai AI Laboratory and Sense Time \citep{2023internlm}, and the GLM \citep{zeng2022glm,du2022glm} to categorize occupations based on their complete set of occupational descriptions. Each of these models operates based on a comprehensive rubric for scoring LLMs exposure. We, then, submit an occupation's description, in conjunction with its title, to each model. In response, each model provides an exposure score. These scores effectively capture whether the time required to complete a task could be halved while maintaining consistent quality, assuming a worker has access to ChatGPT-like LLMs. The Scores are divided into four categories: E0, E1, E2, and E3, the details of each category are presented in Appendix A.

Although we share similar methodologies with \cite{eloundou2023gpts}, several caveats should be noted. First, all the occupational description and prompt are in Chinese, our exercise relies on capabilities of the Chinese large language models. To accomplish better Chinese language performance, we use two top Chinese large language models-GLM and InternLM-in addition to GPT4. Second, the Occupational Classification Dictionary of the People's Republic of China contains only detailed work content and descriptions for each occupation. Therefore, it is infeasible to calculate exposure score at the task level as \cite{eloundou2023gpts}. Instead, we choose to calculate exposure score directly at the occupation level. 

OpenAI has pointed out several weaknesses of the method, such as the validity of the task-based framework, relative versus absolute measures, as well as forward-looking and changing nature of the scores. Another limitation we would like to discuss here is the randomness of LLM scoring. The same prompt can still yields different results from large language models, even with a higher temperature setting. In order to compensate for this issue, we first had each LLM model label each occupation 8 times and calculated the scores. We then took the average as the final scoring result for each LLM model for each occupation.

To compare the consistency of LLM scoring with human scoring, we invited a group of experts in economics and artificial intelligence to serve as judges. We provided them with descriptions of medium-category occupations in China and asked them to score each occupation according to the rating criteria. After collecting the scores from all of the experts, we calculated the average score for each occupation as the final human score. The rating criteria represented the proportion of labor input that large language models could save in each occupation. More details on expert evaluation is presented in Appendix A.

\section{Results and Assessment of Impacts of LLMs on China's Labor Market}

\subsection{Summary Statistics}

We employed the aforementioned methodology to gather results from GPT-4, GLM, and InternLM, subsequently assigning scores to the fine categories occupations. We designated E1 as 1 point, E2 and E3 as 0.5 points each, and E0 as 0 points. It is worth noting that the OCD encompasses 8 large categories, which also include 3 distinctive categories: ``National institutions, party and mass organizations leaders", ``Military personnel", and ``Other personnel not classified elsewhere". In the subsequent sections, our results exclude these three categories of occupations, thereby focusing on 1606 fine-category occupations. The scores assigned to small-category, medium-category and large-category occupations represent the averaged scores of the occupations they encompass from the higher level. A comprehensive summary of these metrics, along with the correlations between different models, is provided in the Table \ref{tab: Summary statistics}.
\begin{table}[htbp]
  \centering
  \caption{Summary Statistics of LLMs Exposure}
    \begin{tabular}{cccc}
    \toprule
    \multicolumn{4}{c}{\textbf{Fine Categories Occupation Level Exposure}} \\
    \midrule
          & \textbf{GLM} & \textbf{InternLM} & \textbf{GPT-4} \\
    \midrule
    count & 1606  & 1606  & 1606  \\
    mean  & 0.44  & 0.18  & 0.24 \\
    std   & 0.26  & 0.18  & 0.21 \\
    \midrule
    \multicolumn{4}{c}{\textbf{Fine Categories Occupation Level Exposure Corr.}} \\
    \midrule
          & \textbf{GLM} & \textbf{InternLM} & \textbf{GPT-4} \\
\cmidrule{2-4}    GLM   & 1.0*** & 0.284*** & 0.1915*** \\
    InternLM & 0.284*** & 1.0*** & 0.2887*** \\
    GPT-4 & 0.1915*** & 0.2887*** & 1.0*** \\
    \midrule
    \multicolumn{4}{c}{\textbf{Medium Categories Occupation Level Exposure}} \\
    \midrule
          & \textbf{GLM} & \textbf{InternLM} & \textbf{GPT-4} \\
    \midrule
    count & 63    & 63    & 63 \\
    mean  & 0.40  & 0.14  & 0.22 \\
    std   & 0.15  & 0.10  & 0.18 \\
    \midrule
    \multicolumn{4}{c}{\textbf{Medium Categories Occupation Level Exposure Corr.}} \\
    \midrule
          & \textbf{GLM} & \textbf{InternLM} & \textbf{GPT-4} \\
\cmidrule{2-4}    GLM   & 1.0*** & 0.5938*** & 0.306* \\
    InternLM & 0.5938*** & 1.0*** & 0.4807*** \\
    GPT-4 & 0.306* & 0.4807*** & 1.0*** \\
    \bottomrule
    \end{tabular}%
  \label{tab: Summary statistics}%
\end{table}%

When examining the scoring across fine-categories occupations, GLM demonstrated an average of 0.44, while InternLM and GPT-4 yielded averages of 0.18 and 0.24 respectively. This suggests that, on average, GLM ascribes stronger capabilities to LLMs and deems occupations to be more susceptible to LLMs exposure. Despite variations in the models' scoring and comprehension of occupations, the scoring outcomes from three models exhibited noteworthy positive correlations. This held true across fine-category, small-category, and medium-category occupations. We synthesized the outcomes from the three models, and throughout the subsequent analysis, unless otherwise specified, we relied on the average scores generated by the three models based on medium-category occupations.

Upon comparing the model scores with the results of human expert scoring in Figure \ref{fig:cor_GLM2+Internlm+GPT_Expert}, we observed substantial positive correlations between the model scores and human expert scores.

\subsection{Occupational Exposure}




Figure \ref{fig:medium_score_GLM2+Internlm+GPT} provides an overview of the medium-category occupation exposure scores by LLMs. It's worth noting that within the highest exposure occupations, such as ``Teachers", ``Economic and financial professionals", and ``Press, publishing and cultural professionals", a strong resonance emerges with recent technological advancements in LLMs, particularly in their capacity to handle a broad spectrum of intricate language-based tasks and even generate functional code based on high-level descriptions of a programming task. Moving to a broader perspective, the large-category occupation exposure scores by LLMs and expert are represented in Appendix B.

The results of our LLMs exposure scores indicate significantly different mechanisms and corresponding impacts on labor market, when compared with previous waves of computer-based automation. Prior research on computer-based automation posited that routine work was most susceptible to replacement by computers \citep{autor2006polarization,acemoglu2011skills}. This ``routinization" hypothesis assumed that ``computers and computer-controlled equipment are highly productive and reliable at performing the tasks that programmers can script - and relatively inept at everything else". The displacement of routine jobs by computers led to a heightened demand for skilled workers in ``abstract" jobs, resulting in wage inequality and job polarization. However, Figure \ref{fig:salary_GLM2+Internlm+GPT} suggests that the labor market repercussions resulting from recent advances in LLMs may diverge. Specifically, this wave of technological change distinguishes itself from earlier waves by potentially substituting numerous tasks within non-routine cognitive analytical jobs that were previously considered immune to automation. In essence, this development suggests a shift in the labor market landscape, where the winners and losers associated with general-purpose technology evolve.

Beyond the distinct impacts on routine and non-routine occupations, LLMs also reshape the experience premium profile. LLMs have the capacity to learn tasks equiring tacit knowledge, knowledge that was once attainable only through experiential learning or learning by doing. As depicted in Figure \ref{fig:wage_growth_GLM2+Internlm+GPT}, LLMs could potentially alter the experience premium and the profile of returns from learning. Our analysis aligns with the insights gleaned from an experimental study \cite{brynjolfsson2023generative}.

\subsection{Composite Occupational Exposure}

In our study, we utilized advanced language models, including GPT-4, GLM, and InternLM, to evaluate the quintet of occupational categories defined by \cite{acemoglu2011skills}. These composite occupational categories are: Non-routine Cognitive (Analytical), Non-routine Cognitive (Interpersonal), Routine Cognitive, Routine Manual, and Non-routine Manual Physical.

To derive the composite exposure scores, the models were tasked with repeatedly scoring various sub-elements within the occupational descriptions. The aggregate score for each composite occupation was then computed as the mean of these evaluations.

The analytical results, illustrated in Figure \ref{fig: skills and llms}, reveal that cognitive tasks, particularly those of a non-routine nature, are the most significantly impacted by large language models (LLMs). Routine manual tasks are also expected to experience considerable influence, while tasks classified as Interpersonal and Non-routine Manual Physical are less affected by these advanced models.

This differential impact underscores the nuanced effects of LLMs across task categories. Notably, our findings diverge from the “routinization hypothesis” \citep{autor2003skill}, which posits that information and communication technologies (ICT) primarily automate “routine” codifiable tasks traditionally performed by middle-wage workers. In Section 5, we develop a theoretical model to further investigate the mechanisms driving the different occupational exposures of traditional technologies and LLMs.

\subsection{Industrial and Demographic Exposure}

As highlighted in the preceding Data section, we leveraged the 2016 CLDS dataset, which encompasses comprehensive industry and occupational information for each individual in employment. Utilizing this dataset, we derived the occupational distribution within each industry. This distribution, when coupled with the exposure scores attributed to the corresponding occupations, enabled the computation of a weighted occupational intensity index for each industry. 

Figure \ref{fig: ind_score_GLM2+Internlm+GPT} provides insights into substantial variations in the exposure levels of LLMs across distinct industries. Notably, sectors with the highest scores, including education, culture and arts, radio, film and television, exhibit heightened susceptibility. This observation underscores the pronounced influence of the current technological landscape on these domains. Conversely, industries with lower scores, such as agriculture, forestry, husbandry, fisheries, mining, manufacturing and construction- persist in displaying a range of distinctive complexities. Despite the overarching advancements in artificial intelligence across diverse domains, these sectors present enduring challenges. Their intricate nuances, customized requirements, and the enduring necessity for human judgment contribute to their current resilience against full automation and replacement in the foreseeable future.

Armed with industrial exposure data, we proceed to compute demographic exposure by integrating it with the demographic distribution across industries. Relying on the tabulations of census data from 2020 in China, we observe a notable concentration of  youth employment within the tertiary sector. To illustrate, during 2020, 42.5\% of individuals aged 16-19 and 34.9\% of those aged 20-24 were engaged in consumer service industries. Figure \ref{fig: demographic exposure} showcases our demographic exposure scores, revealing that LLMs wield a more pronounced influence on the employment prospects of young individuals.


\subsection{Vacancies}

Figure \ref{fig: share of vacancy llms} and \ref{fig: percentage change of share of vacancy llms} offer a comprehensive overview of potential impacts on future labor demand whinin China. The presence of a positive correlation between occupational exposure and vacancy share, as depicted in Figure \ref{fig: share of vacancy llms}, suggests that occupations with a higher vacancy share may potentially experience more pronounced disruptive effects from LLMs in the future. Contrary to the prevailing assumptions, the occupational structure in China tends to exacerbates rather than ameliorate the disruption caused by LLMs. Figure \ref{fig: percentage change of share of vacancy llms} unveils an additional layer of concern related to the disruptive effects of LLMs. The positive correlation between occupational exposure and growth rate of vacancy share implies that the advancements of LLMs could unexpectedly reverse the prevailing labor demand trend, thereby intensifying the structural unemployment predicament in the future.

\section{The Model}
This section introduces a theoretical framework designed to enhance
understanding of task automation and its relationship with large language
models (LLMs). Drawing on prior empirical findings, we have discerned a
predominant impact of automation on occupations involving non-routine
cognitive tasks. This observation marks a departure from earlier
conclusions, such as those posited by \cite{autor2003skill}, which suggested
that automation primarily affects occupations characterized by routine
tasks. A key insight from our analysis is the pronounced effect of
automation on occupations heavily reliant on non-routine cognitive
analytical tasks, in contrast to the relatively minimal impact on manual
physical tasks. Occupations involving interpersonal tasks demonstrate a
moderate degree of susceptibility to LLMs, positioned between cognitive and
manual tasks in terms of exposure. Additionally, our findings indicate a
correlation between occupations with higher wage and experience premiums and
their increased vulnerability to LLMs. This aligns with the recent research
by \cite{eloundou2023gpts} and \cite{eisfeldt2023generative}, who have identified
that technological advancements disproportionately affect workers in the
higher echelons of the wage spectrum. To elucidate the varied patterns of
occupational exposure to different technological advancements, our proposed
theoretical model offers a nuanced perspective on the distinct mechanisms
driving occupational automation. This model aims to bridge the gap in
understanding the differential impacts of technology across various
occupations.

The central thesis of our analysis hinges on the distinct technological
attributes inherent to the traditional AI, deep learning AI, and LLMs AI. Essentially, our theoretical model posits that the unique
patterns of occupational exposure observed are directly attributable to the
underlying technological principles governing automation processes. To
effectively articulate this proposition, we integrate the principles of
information theory into our task framework. This integration facilitates a
more comprehensive understanding of the disparities between various AI technological paradigms, particularly highlighting the differences in
productivity between models based on traditional AI and those developed using deep learning AI and LLMs AI technology.

The structure of the theoretical section of our study is methodically
organized as follows: Subsection 1 delineates the environments and
competitive equilibrium within a stylized task model. This foundational
subsection lays the groundwork for understanding the basic dynamics of task
allocation and performance. Subsection 3 seamlessly integrates traditional
AI technology into our task framework. This integration is
instrumental in reexamining and substantiating the routinization hypothesis,
which asserts that traditional AI technologies are predominantly
applied to routine tasks. Subsection 4 is pivotal as it illuminates the
distinct patterns of occupational exposure. It accomplishes this by
elucidating the scaling laws characteristic of deep learning AI and LLMs AI technology. This
subsection is critical in contrasting the logic of these advanced
technologies with that of traditional AI, thereby
offering a nuanced perspective on how technological advancements influence
various occupational sectors differently.

\subsection{The Benchmark Model}

The benchmark model shares common basic settings with the other models, but
it lacks task automation originated from the different technologies.

\subsubsection{The Environment}

The unique final good is produced by combining a continuum of tasks
represented by the unit interval $\left[ 0,1\right] $. In particular,%
\begin{equation}
Y=\exp \left[ \int_{0}^{1}\ln y\left( i\right) di\right] ,  \label{1-1}
\end{equation}%
where $Y$ denotes the output of a unique final good and $y\left( i\right) $
is referred to as the output of task $i$. We assume all markets to be
competitive. Throughout, we choose the price of the final good as the
numeraire.

Each task is produced by the following technology function%
\begin{equation}
y\left( i\right) =A_{L}\alpha _{L}\left( i\right) l\left( i\right)
+A_{H}\alpha _{H}\left( i\right) h\left( i\right) +A_{M}\alpha _{M}\left(
i\right) m\left( i\right) ,  \label{1-2}
\end{equation}%
where $A$ terms represent a factor-augmenting technology, and $\alpha $
terms are the task productivity profiles, designating the productivity of
low or high skill workers and models in different tasks. It is important to
see that this production function for task services implies that each task
can be performed by low or high skill workers and a technological model, but
the comparative advantage of skill groups and different model differs across
tasks which is captured by the $\alpha $ terms. The differences in
comparative advantage play a central role in a Ricardian model of the labor
market.

In our analysis, the deliberate use of the term ``model" instead of ``capital"
accentuates our focus on the intangible aspects of AI automation technologies.
These models are distinguished by their unique technological logics. To
deepen our understanding of the varying technological logics underpinning
these automation technologies, we delve into the prominent discourse on
``Software 1.0 vs 2.0", notably advanced by Andrej Karpathy, Sr. Director of
AI at Tesla, in his 2017 Medium post titled ``Software 2.0".

Software 1.0 represents the paradigm prevalent in traditional software
development, employing languages such as Python, C++, among others. In this
paradigm, programmers craft explicit instructions in the form of code, which
the software executes. Each line of code is a directive that specifies a
certain aspect of the program, guiding it towards a desired behavior.

Karpathy posits that the advent of neural networks, which are foundational
to modern deep learning AI technology, heralds a paradigmatic shift in software
development. According to him, Software 2.0 centers on compiling a dataset
that defines the ``desirable behavior" or the problem to be solved. This is
followed by establishing the basic structure or ``skeleton code" of the
neural network's architecture. The network is then allowed to optimize
itself for the solution. Crucially, Software 2.0 diverges from Software 1.0
in that it doesn't involve code written by programmers in the traditional
sense. Instead, the ``code" emerges from computations performed on large
datasets.

This insightful discussion about the technological distinctions between traditional AI
and deep learning AI or LLMs AI informs our categorization of models'
technologies into three paradigms: traditional AI, deep learning AI, and the LLMs AI.

Our theoretical framework also incorporates two pivotal concepts from
information theory. The first is entropy, which aids in defining the
complexity inherent in each occupation. The second concept is the
Kullback-Leibler divergence (KL-divergence), which is instrumental in
specifying the productivity of different models. These concepts are critical
for a nuanced understanding of the impact of various technological paradigms
on different occupations.

Let us utilize concept of entropy to define complexity of different tasks
and comparative advantage of skills. Let $X$ be a discrete random variable
with probability mass function $P(x)$. The entropy (or Shannon entropy) of $%
X $ is
\begin{equation}
H(P)=E_{P}\left[ \log \frac{1}{P(X)}\right] =\sum_{x}P(x)\log \frac{1}{P(x)}.
\label{Def_entropy}
\end{equation}%
Entropy is a quantification of uncertainty in a distribution. The higher the
entropy of the distribution, the more information is contained in the
realization of a random variable it governs. Let us assume that there exists
a distribution function $P_{i}$ which characterize the relevant information
in task $i$. Therefore, the complexity of a task $i$ is defined by entropy $%
H_{i}\left( P\right) $. Furthermore, we assume that skilled worker has a
comparative advantage in the more complex tasks. Specifically, we assume
that $H_{i}$ is evenly distributed on the domain $\left[ 0,\bar{H}\right] $
and $\bar{H}=\max \{H_{i}\}$. Now we can index task $i$ by their
correspondent relative complexity $H_{i}/\bar{H}$.

\textbf{Assumption} 1 $\alpha _{L}\left( i\right) =\left( 1-H_{i}/\bar{H}%
\right) ^{\eta }=\left( 1-i\right) ^{\eta }$ and $\alpha _{H}\left( i\right)
=\left( H_{i}/\bar{H}\right) ^{\eta }=i^{\eta }$, $\eta >0$.

The assumption 1 characterize the comparative advantage of skills across
tasks. The implication of the assumption is that compared to unskilled
workers skilled worker have a competitive edge on the complex tasks.

\subsubsection{Competitive Equilibrium without Task Automation}

We will characterize the competitive equilibrium without task automation in
the following. There will exist a threshold $I$ such that all tasks $i<I$
will be performed by low skill workers, and all tasks $i>I$ will be
performed by high skill workers.

\textbf{Lemma 1}: In equilibrium, there exists a threshold $I$ such that $%
0\leq I\leq 1$ and for any $0\leq i<I$, $h\left( i\right) =0$, for any $%
I<i\leq 1$, $l\left( i\right) =0$.

The competitive equilibrium can be characterized by the following four
conditions:

\begin{enumerate}
\item Law of one price for skills%
\begin{equation}
W_{L}=p\left( i\right) A_{L}\alpha _{L}\left( i\right) \ \text{\ for any }%
i<I,  \label{wl}
\end{equation}%
\begin{equation}
W_{H}=p\left( i\right) A_{H}\alpha _{H}\left( i\right) \ \text{\ for any }%
i>I.  \label{wh}
\end{equation}

\item Goods market (or task market) clearing condition demands:%
\begin{equation}
p\left( i\right) y\left( i\right) =p\left( i^{\prime }\right) y\left(
i^{\prime }\right) =Y.  \label{GC}
\end{equation}

In the interval $i<I$, the following equation is satisfied
\[
p\left( i\right) A_{L}\alpha_{L}\left( i\right) l\left( i\right) =p\left(
i^{\prime}\right) A_{L}\alpha_{L}\left( i^{\prime}\right) l\left(
i^{\prime}\right) .
\]

Combining with the condition of the law of one price for skills, we obtain

\begin{equation}
l\left( i\right) =l\left( i^{\prime}\right) \ \text{for any }i<I,
\label{1-3}
\end{equation}

and%
\begin{equation}
h\left( i\right) =h\left( i^{\prime }\right) \text{ for any }i>I.
\label{1-4}
\end{equation}

\item No arbitrage condition across skills: the threshold task $I$ can be
profitably produced using either skilled or unskilled workers:%
\begin{equation}
A_{L}\alpha_{L}\left( I\right) l\left( I\right) =A_{H}\alpha_{H}\left(
I\right) h\left( I\right) .  \label{1-5}
\end{equation}

\item Labor market clearing condition:%
\begin{equation}
\int_{0}^{I}l\left( i\right) di=L,  \label{1-6}
\end{equation}%
\begin{equation}
\int_{I}^{1}h\left( i\right) di=H.  \label{1-7}
\end{equation}
\end{enumerate}

By (\ref{1-3}), (\ref{1-4}), (\ref{1-6}), and (\ref{1-7}), we obtain that
\begin{equation}
l\left( I\right) =L/I,  \label{1-8}
\end{equation}%
\begin{equation}
h\left( I\right) =H/\left( 1-I\right) .  \label{1-9}
\end{equation}

Plugging (\ref{1-8}), and (\ref{1-9}) into (\ref{1-5}), we can obtain
\begin{equation}
I=1/\left( 1+\left( \frac{A_{H}H}{A_{L}L}\right) ^{\frac{1}{1+\eta }}\right)
.  \label{I}
\end{equation}%
Wage difference is determined by the productivity difference at the threshold%
\[
\frac{W_{H}}{W_{L}}=\frac{A_{H}\alpha _{H}\left( I\right) }{A_{L}\alpha
_{L}\left( I\right) }=\left( \frac{A_{H}}{A_{L}}\right) ^{1/\left( 1+\eta
\right) }\left( \frac{H}{L}\right) ^{-\eta /\left( 1+\eta \right) }\ .
\]%
The following proposition summarizes the equilibrium allocation and
corresponding wage premium.

\textbf{Proposition 1}: The competitive equilibrium is characterized by a
threshold condition $I$%
\[
I=1/\left( 1+\left( \frac{A_{H}H}{A_{L}L}\right) ^{\frac{1}{1+\eta }}\right)
,
\]%
and labor allocations across occupations. In the domain $i\in \left[ 0,I%
\right] $, $L/I$ measure of unskilled workers are allocated to each task,
meanwhile, In the domain $i\in \left[ I,1\right] $, $H/\left( 1-I\right) $
measure of skilled workers are allocated to each task.

The corresponding wage premium is

\begin{equation}
\frac{W_{H}}{W_{L}}=\left( \frac{A_{H}}{A_{L}}\right) ^{1/\left( 1+\eta
\right) }\left( \frac{H}{L}\right) ^{-\eta /\left( 1+\eta \right) }.
\label{W_pre_1}
\end{equation}%
The two terms on the right hand of previous equation (\ref{W_pre_1})
correspond to two forces in Tinbergen's race: technology and education. The
terms $\left( \frac{A_{H}}{A_{L}}\right) ^{1/\left( 1+\eta \right) }$
represents the impacts of the skill-biased technology change on the relative
demand of skills, otherwise the terms $\left( \frac{H}{L}\right) ^{-\eta
/\left( 1+\eta \right) }$ embodies the impacts of the relative supply of
skills which is determined by education.

\subsection{The Model with Different AI Technologies}

In this subsection, let us introduce the automation model with different AI technologies to perform a task. 

\subsubsection{The Comparative Advantage of Traditional AI}

The primary innovation of our theoretical construct is the nuanced
elucidation of the comparative productivity advantage of various models in
specific tasks, as informed by information theory. This study systematically
considers models grounded in three technological paradigms: traditional
AI, deep learning AI, and large
language models (LLMs) AI. Each model, underpinned by these
divergent technologies, has been rigorously trained to manifest optimal
task-specific behavior, achieving a level of precision that renders any
distinction between the model's performance and an idealized version of task
execution virtually undetectable. Consequently, the accuracy of a model's
performance is directly proportional to its productivity in automated tasks. Hereafter, we denote all endogenous variable in this economy with an upper dagger.

A fundamental postulate of this research is that the inability to
distinguish between a model's output and the theoretically perfect execution
of a task denotes competence. This principle is a cornerstone in the
behavioral evaluation of machine intelligence, epitomized by seminal tests
such as Turing's Imitation Game.

Specifically, we resort to the foundational concept in information theory.
KL-divergence, also known as relative entropy, quantifies how different two
distributions $P$ and $Q$ are. It is arguably a central pillar of
information theory. The KL-divergence between $P$ and $Q$ is
\begin{equation}
D\left( P||Q\right) =\sum_{x}P(x)\log \left( \frac{P(x)}{Q(x)}\right) .
\label{Def_KL}
\end{equation}%
Intuitively, larger log likelihood ratios $\log \left( \frac{P(x)}{Q(x)}%
\right) $ reflect distributions that are more different. KL-divergence
aggregates these log likelihood ratios by weighting them with respect to
their probabilities under a reference distribution. Now we can define a
model's KL-divergence $D_{i}$ where $P_{i}$ is the information distribution
embodied in a model performing task $i$ and $Q_{i}$ is the information
distribution that a best model can achieve.

Now we can resort to we KL-divergence to characterize the productivity
profile $\alpha _{M}\left( i\right) $. Let assume that for each task, we
have a productivity upper bound $\bar{A}$, and productivity in task $i$ for
a model with traditional AI is%
\[
\alpha _{M}\left( i\right) =F\left( \tilde{D}_{i}\right) =\left( 1-\tilde{D}%
\left( i\right) /\bar{D}\right) ^{\mu },\ A_{m}=\bar{A},
\]%
Let us assume that $\tilde{D}_{i}$ is evenly distributed on the domain $%
\left[ 0,\bar{D}\right] $ and $\bar{D}=\max \{\tilde{D}_{i}\}$, therefore,
\[
\alpha _{M}\left( i\right) =\left( 1-i\right) ^{\mu },
\]%
which implies that the model tends to have a higher productivity in less
complex tasks. Furthermore, to ensure an internal solution, we assume that
\[
\bar{A}>A_{L},\ \mu >\eta .
\]%
We also impose a exogenous price of models $M$ as $F$.

The aforementioned assumption elucidates the underlying technical rationale
inherent in traditional AI. Drawing an analogy with the
paradigm of software 1.0, this technology facilitates task automation by
encoding explicit rules pertinent to a specific task. It is evident that the
process of encoding rules is considerably more straightforward in tasks of
lesser complexity. In contrast, for more complex tasks where rules are
either excessively intricate or ambiguously defined, the codification
process becomes substantially more challenging. Hence, our postulation
aligns with the routinization hypothesis, which posits that traditional
AI models demonstrate a comparative advantage in automating routine
or less complex tasks. Consequently, in line with these assumptions, our
analysis reaffirms the routinization hypothesis, indicating a tendency for
models based on traditional AI to predominantly automate
routine tasks.

\textbf{Proposition 2}: In competitive equilibrium, there exists two
threshold $\tilde{I}_{1}$ and $\tilde{I}_{2}$ such that $0<\tilde{I}_{1}<%
\tilde{I}_{2}<1$ and for any $\leq i<\tilde{I}_{1}$, $\tilde{l}\left(
i\right) =\tilde{h}\left( i\right) =0$, for any $\tilde{I}_{1}<i<\tilde{I}%
_{1}$, $\tilde{h}\left( i\right) =\tilde{m}\left( i\right) =0$, for any $%
\tilde{I}_{2}<i\leq 1$, $\tilde{l}\left( i\right) =\tilde{m}\left( i\right)
=0$. Furthermore, the wage premium increases.

Proof see appendix.

Proposition 2 in our study reveals that models leveraging traditional AI technology are inclined to replace routine tasks, typically
executed by unskilled workers. This leads to a reduction in the task domain
available to unskilled workers, subsequently widening the wage disparity
between skilled and unskilled labor.

\subsubsection{The Comparative Advantage of Deep Learning AI and LLMs AI Technology}

In this subsection, let us introduce the automation model with deep learning AI and LLMs AI technology to perform a task. We use the hat operator to indicate endogenous variables in this economy.

At present, in the realm of artificial intelligence, Large Language Models
(LLMs) represent the most advantageous form of generative artificial
intelligence. These AI systems, designed to predict subsequent words based
on preceding text, are further refined through fine-tuning to align with
human instructions and preferences. LLMs are distinguished by their
foundation in deep neural networks, boasting an expansive range of billions,
or even trillions, of parameters in cutting-edge models. This positions both deep learning
AI and LLM AI models firmly within the realm of the "Software 2.0" paradigm,
characterized by a shift towards data-driven learning and adaptive
algorithms.

The advent of transformer models, as introduced by \citep{vaswani2017attention},
marks a pivotal development in the evolution of LLMs. Transformers
incorporate an "attention mechanism" that dynamically assigns varying levels
of significance to different words within a text. This innovation
significantly enhances the model's ability to decipher complex patterns and
dependencies in language, thereby improving the efficiency and
interpretative capabilities of language models.

Now let us assume that deep learning AI and LLMs AI models represent a distinct technological
logic. Still we assume that KL-divergence is a measurement of productivity
loss. Let assume that for each task, we have a productivity upper bound $%
\bar{A}$, and productivity in task $i$ for an AI model are%
\[
\alpha _{M}\left( i\right) =F\left( \hat{D}\left( i\right) \right) =\left( 1-%
\hat{D}\left( i\right) \right) ^{\eta },\ A_{m}=\bar{A},
\]%
where%
\begin{equation}
\hat{D}\left( i\right) =L\left( N,G\right) =C+\frac{a}{N\left( i\right)
^{\alpha }}+\frac{b}{G^{\beta }}.  \label{scaling}
\end{equation}

Let us assume there is a different data accumulation pattern across task.
Now assume that $N\left( i\right) =\bar{N}$, if $i\in \left[ \underline{I},%
\bar{I}\right] $, and $I<\underline{I}<\bar{I}<1$ and $N\left( i\right) =0$,
if $i\notin \left[ \underline{I},\bar{I}\right] $. A key conceptual
contribution from the field of AI, particularly relevant to this discussion,
is encapsulated in (\ref{scaling}), which represents a well-established
empirical regularity known as the scaling law. This concept, highlighted in
recent AI literature by researchers such as \citep{kaplan2020scaling}, \citep{hoffmann2022training}, and \citep{barnett2023scaling}, articulates a parametric
form for Kullback-Leibler divergence. This divergence is expressed as a
power law in terms of data ($N$) and parameter count ($G$), with $C$
denoting the irreducible loss, and $a$, $b$, $\alpha $, and $\beta $ are
constants. The robustness of these relationships is underscored by their
persistence across various orders of magnitude. Furthermore, power-law
scaling laws have been identified as a ubiquitous presence in the deep learning AI and LLMs AI domain,
with their application extending to areas like auto-regressive modeling and
reinforcement learning, and have been the subject of extensive theoretical
investigation.

Given that deep learning AI and LLMs AI models exhibit similar power scaling laws, our focus
can be directed specifically towards LLMs AI without a significant loss of
generality. In the ensuing analysis, we conceptualize technological changes
in LLMs as a substantial increase in the parameter count $G$, facilitating
the execution of tasks within the specified domain $\left[ \underline{I},%
\bar{I}\right] $ by LLMs. This approach allows for a nuanced understanding
of the impact and potential of LLMs AI in transforming various task
domains.

\subsection{Discussions}
In this subsection, we discuss our model with specific AI Applications. And we also elaborate theoretical and policy implications from theoretical model furthermore.

\subsubsection{Discussions With AI Applications}
The healthcare industry has long been a dynamic testing ground for cutting-edge AI capabilities. Recent advances in large language models (LLMs) have enabled these systems to acquire broader competencies, allowing for application across a wide range of healthcare tasks. Given the pervasive use of text in healthcare practices, healthcare-specific LLMs, trained on extensive medical language data, have demonstrated exceptional performance in processing medical texts, conducting dialogues, making diagnoses, and supporting education and consultation. We focus here on two particular applications: medical diagnosis and report generation.

LLMs have been employed to predict the most likely diagnoses based on medical tests and patients' personal descriptions. These models have proven effective in enhancing diagnostic processes, showing strong generalist abilities across various diseases. Despite their diagnostic potential, LLMs still face significant challenges. The lack of transparency—where clinicians are unable to fully trace the reasoning behind a model's decisions—remains a key concern in medical practice.

LLMs have also demonstrated remarkable potential in generating medical reports, such as radiology reports, discharge summaries, and referral letters. These models excel at synthesizing information from a variety of sources, including electronic health records (EHRs), medical literature, and clinical guidelines, to produce coherent, informative reports. Medical report generation is often a tedious and time-consuming task for doctors; thus, the use of LLMs can significantly alleviate their workload.

While these applications confirm that both medical diagnosis and report generation require highly skilled expertise, LLMs are showing the potential to empower a broader group of workers to perform these expert tasks. As AI continues to evolve, it could ultimately complement and enhance the judgment of professionals, broadening the scope of tasks that non-experts can manage. This trend is applicable beyond healthcare, suggesting that LLMs can enable more workers to perform complex, high-stakes tasks by augmenting their skills and supporting their professional judgment.

Our theoretical model aligns with this vision. We predict that LLMs will reshape the value and nature of human expertise, helping to narrow the income inequality between skill levels. Several recent empirical studies corroborate this perspective. For example, a controlled experiment by \cite{peng2023impact} found that GitHub Copilot, a generative AI programming aid, significantly boosted productivity: the group using Copilot completed programming tasks approximately 56\% faster than the control group without access to the tool. Similarly, a study by \cite{noy2023experimental} found that access to ChatGPT improved the speed and quality of writing tasks among professionals such as marketers, grant writers, consultants, and managers. The ChatGPT group reduced the time spent on tasks by 40\%, and even the least skilled writers in this group performed as well as the median writer in the control group—highlighting a significant boost in writing quality. Moreover, ChatGPT allowed the most capable writers to work faster and enabled the less capable to not only write faster but also to improve the quality of their output. This narrowing of the productivity gap underscores the potential of LLMs to democratize expertise.

To understand the fundamental difference between traditional AI and modern deep learning-based LLMs, we need to focus on the underlying technological logic. Both AI models and machines require software code to execute tasks. In traditional AI, this code follows a rule-based structure: developers must explicitly define each step in the process, particularly for tasks that are repetitive or relatively simple. This makes routine, less complex tasks more susceptible to automation.

In contrast, deep learning AI and LLMs follow a data-driven methodology. Developers no longer need to predefine the rules; instead, they feed vast datasets into an algorithm, which then learns to perform tasks. For instance, ChatGPT doesn't rely on specific, preset rules for generating poetry; it learns this skill through exposure to an enormous amount of language data. This data-driven approach enables deep learning AI and LLMs to tackle tasks that lack clear, codifiable rules—typically found in more complex, skilled domains. Consequently, LLMs are more suited to automating tasks traditionally performed by skilled workers, rather than simple, rule-based jobs.

\subsubsection{Theoretical and Policy Implications}

Our model incorporates entropy-based information theory into a stylized task-based framework. This task-based model is well-suited to characterizing comparative advantages. To analyze the comparative advantage of skills and technologies, we need a consistent way to define the productivity profiles of skills and technologies across occupations. Specifically, we characterize task complexity using the concept of entropy and define the relative productivity of skills across occupations with varying complexity levels.

Furthermore, the introduction of KL-divergence allows us to model the relative efficiency of AI models across occupations. The central idea is that cognitive tasks, at their core, involve making predictions—such as forecasting GDP growth, predicting stock prices, or generating responses in the case of LLMs. Intuitively, the productivity of cognitive tasks is determined by prediction accuracy, which we represent through KL- divergence. KL-divergence is also used in the ``neural scaling laws" to measure an AI model’s prediction accuracy, which gives task performance a solid foundation by directly borrowing well-established AI technological principles and bridging them with information theory.

Our theoretical framework uniquely integrates information theory into a
stylized task model. This integration accomplishes dual objectives. Firstly,
it enables the application of concepts like entropy and KL-divergence to
quantify the complexity of tasks and the efficacy of algorithms in specific
tasks. Secondly, information theory acts as a conduit, linking economic
models with the body of literature in computer science. This synthesis
provides a robust micro-foundation for the understanding of deep learning AI and LLMs AI technology.

In our theoretical model, the key distinction between traditional AI and LLMs as automation technologies lies in their operational logic: traditional AI depends on pre-programmed rules to perform tasks, while LLMs leverage neural scaling laws and data-driven methods. This fundamental difference in technological logic results in varying comparative advantages across tasks. Traditional AI excels at performing less complex tasks typically carried out by unskilled workers. The automation of such tasks increases wage premiums between skilled and unskilled workers by eroding the position of unskilled workers.

In contrast, LLMs, powered by neural scaling laws and vast amounts of data and computational resources, exhibit a comparative advantage in data-intensive and complex tasks, usually performed by skilled workers. This shift erodes the expertise position of skilled workers and leads to a more flattened wage distribution between skill levels.

The overarching insight from our theoretical model is that algorithms, based
on distinct technological logic, are predisposed to disrupt different types
of tasks. These variances in technological underpinnings will inevitably
result in divergent outcomes in terms of productivity and income
distribution. This distinction is why the progression of LLMs AI technology
will generate outcomes that deviate from those predicted by the
routinization hypothesis. Fundamentally, the influence of LLMs AI technology on
labor markets is anticipated to be distinctly different from that of
traditional AI. In essence, this era of technological
advancement represents a departure from past trends.

\section{Discussions and Conclusions\label{sec:Conclusion}}

The utopian vision of the information age anticipated that computers would democratize information and flatten economic hierarchies. However, this vision has not materialized; instead, the opposite has occurred. Labor income inequality has widened significantly, with computerization facilitating the unprecedented concentration of decision-making power among elite experts, who now have access to abundant and inexpensive information \citep{deming2021growing}. Our theoretical model predicts that artificial intelligence, by leveraging vast datasets and computational power to synthesize information and rules, could enable a broad range of workers equipped with foundational training to perform high-stakes decision-making tasks. These tasks, traditionally monopolized by elite experts—such as doctors, lawyers, software engineers, and college professors—may become accessible to more individuals. In this regard, our theoretical perspective suggests that artificial intelligence has the potential to reduce income inequality, contrary to the trends observed with earlier technologies.

\cite{ge2014changes} documented that between 1992 and 2007, during a period of rapid economic growth in China, the average real wage increased by 202\%, accompanied by a sharp rise in wage inequality. While job polarization did not occur in China, the wages of elite experts rose dramatically, indicating that information technology has played a significant role in exacerbating income inequality. Our theoretical model applies to this context as well, reflecting the mechanisms by which technological advancements affect wage distribution. These results align with the broader discussions on technological impacts, as noted by \cite{autor2024applying}. To summarize, our theoretical model aims to interpret the differing implications of various technologies on income inequality based on their distinct features. In this sense, we believe the model captures at least part of the momentum shaping labor market demand in China.

In terms of policy implications, our theoretical findings for artificial intelligence should be viewed as scenario analyses rather than definitive predictions. The impact of any technology depends not only on its inherent features but also on institutional responses. The first policy implication of our model is the disruptive impact of advanced AI technologies, such as LLMs, on expert labor. AI has the potential to reduce scarcity by empowering more workers to perform expert-level tasks. However, this potential will not be realized automatically. For example, a recent experiment found that providing AI assistance to radiologists did not improve diagnostic quality, even though AI predictions were at least as accurate as those of two-thirds of the doctors studied \citep{agarwal2023combining}. This outcome occurred because doctors often overrode AI predictions, highlighting their lack of understanding of how to use the tools effectively. Thus, a key policy orientation for the future is to develop training programs that enable workers to effectively utilize AI tools.

The second policy implication concerns the substantial uncertainties surrounding the technological impacts on labor markets, as evidenced by the unrealized utopian vision of the information age. To design effective policy tools, it is crucial to systematically track labor market demand conditions. Unlike the U.S., which has a system like ONET, China currently lacks a comprehensive statistical framework to monitor occupational demand. Establishing such a system should be a priority, given the impending disruptive waves of AI technologies.

As a general-purpose technology, the introduction and proliferation of large language models constitute a substantial technological upheaval with significant implications for the overall economy. This paper employs measures of occupational exposures to LLMs, in conjunction with aggregate assessments of occupational composition, to evaluate the potential impact of LLMs on labor market within China's economy. Our findings suggest that the potential effect of the release of LLMs could have a substantial effect on the Chinese labor market, potentially altering previous labor demand trends. LLMs indeed represent a noteworthy shock to the Chinese labor market. Furthermore, we develop a theoretical model to provide a deeper understanding of the reasons behind occupational exposure to LLMs, which diverges from the prediction of the routinization hypothesis.

\newpage
\bibliographystyle{apalike}
\bibliography{reference}

\newpage
\begin{appendices}

\setcounter{equation}{0}
\numberwithin{equation}{section}

\renewcommand{\thesection}{Appendix A}
\section{Proofs}\label{appendix_a}
\renewcommand{\thesection}{A}

\ 
\newline
The appendix A contains the algebra and proofs for theoretical model.

\subsection{Proof of the Model with Traditional AI}

Now we know the task specific productivity profiles are%
\[
\alpha _{L}\left( i\right) =\left( 1-i\right) ^{\eta },
\]%
\[
\alpha _{H}\left( i\right) =i^{\eta },
\]%
\[
\alpha _{M}\left( i\right) =\left( 1-i\right) ^{\mu }.
\]

The competitive equilibrium can be characterized by the following conditions:

\begin{enumerate}
\item Law of one price for skills%
\[
F=\tilde{p}\left( i\right) A_{M}\alpha _{M}\left( i\right) \text{ for any }i<%
\tilde{I}_{1},
\]%
\[
\tilde{W}_{L}=\tilde{p}\left( i\right) A_{L}\alpha _{L}\left( i\right) \
\text{\ for any }\tilde{I}_{2}>i>\tilde{I}_{1},
\]%
\[
\tilde{W}_{H}=\tilde{p}\left( i\right) A_{H}\alpha _{H}\left( i\right) \
\text{\ for any }i>\tilde{I}_{2}.
\]

\item Goods market (or task market) clearing condition demands:%
\[
\tilde{p}\left( i\right) \tilde{y}\left( i\right) =\tilde{p}\left( i^{\prime
}\right) \tilde{y}\left( i^{\prime }\right) =\tilde{Y}.
\]

In the three intervals, the following equations are satisfied%
\[
\tilde{p}\left( i\right) A_{L}\alpha _{L}\left( i\right) \tilde{l}\left(
i\right) =\tilde{p}\left( i^{\prime }\right) A_{L}\alpha _{L}\left(
i^{\prime }\right) \tilde{l}\left( i^{\prime }\right) ,
\]%
\[
\tilde{p}\left( i\right) A_{H}\alpha _{H}\left( i\right) \tilde{h}\left(
i\right) =\tilde{p}\left( i^{\prime }\right) A_{H}\alpha _{H}\left(
i^{\prime }\right) \tilde{h}\left( i^{\prime }\right) ,
\]%
\[
\tilde{p}\left( i\right) A_{M}\alpha _{M}\left( i\right) \tilde{m}\left(
i\right) =\tilde{p}\left( i^{\prime }\right) A_{M}\alpha _{M}\left(
i^{\prime }\right) \tilde{m}\left( i^{\prime }\right) .
\]

Combining with the condition of the law of one price for skills, we obtain%
\[
\tilde{m}\left( i\right) =\tilde{m}\left( i^{\prime }\right) \text{ for any }%
i<\tilde{I}_{1},
\]%
\[
\tilde{l}\left( i\right) =\tilde{l}\left( i^{\prime }\right) \text{ for any }%
\tilde{I}_{2}>i>\tilde{I}_{1},
\]

similarly%
\[
\tilde{h}\left( i\right) =\tilde{h}\left( i^{\prime }\right) \text{ for any }%
i>\tilde{I}_{2}.
\]

\item No arbitrage condition across skills and models: the threshold task $%
\tilde{I}_{1}$ can be profitably produced using either models or unskilled
workers, and the threshold task $\tilde{I}_{2}$ can be profitably produced
using either skilled or unskilled workers,
\[
A_{L}\alpha _{L}\left( \tilde{I}_{1}\right) \tilde{l}\left( \tilde{I}%
_{1}\right) =A_{M}\alpha _{M}\left( \tilde{I}_{1}\right) \tilde{m}\left(
\tilde{I}_{1}\right) ,
\]%
\[
A_{L}\alpha _{L}\left( \tilde{I}_{2}\right) \tilde{l}\left( \tilde{I}%
_{2}\right) =A_{H}\alpha _{H}\left( \tilde{I}_{2}\right) \tilde{h}\left(
\tilde{I}_{2}\right) .
\]

\item Labor market clearing condition:%
\[
\int_{0}^{I}\tilde{l}\left( i\right) di=L,
\]%
\[
\int_{I}^{1}\tilde{h}\left( i\right) di=H.
\]
\end{enumerate}

With previous conditions, we obtain that
\[
\tilde{l}\left( i\right) =L/\left( \tilde{I}_{2}-\tilde{I}_{1}\right) ,
\]%
\[
\tilde{h}\left( i\right) =H/\left( 1-\tilde{I}_{2}\right) .
\]

We can prove the proposition 2 by contradiction. If $\tilde{I}_{2}\leq I$,
we know that
\[
H/\left( 1-\tilde{I}_{2}\right) \leq H/\left( 1-I\right) ,
\]%
and%
\[
L/\left( \tilde{I}_{2}-\tilde{I}_{1}\right) >L/I.
\]%
With the No arbitrage condition at the threshold, we know that
\[
A_{L}\alpha _{L}\left( \tilde{I}_{2}\right) \tilde{l}\left( \tilde{I}%
_{2}\right) =A_{H}\alpha _{H}\left( \tilde{I}_{2}\right) \tilde{h}\left(
\tilde{I}_{2}\right) ,
\]%
i.e.%
\begin{equation}
\frac{A_{L}}{A_{H}}=\frac{\alpha _{H}\left( \tilde{I}_{2}\right) }{\alpha
_{L}\left( \tilde{I}_{2}\right) }\frac{H/\left( 1-\tilde{I}_{2}\right) }{%
L/\left( \tilde{I}_{2}-\tilde{I}_{1}\right) }  \label{c_1}
\end{equation}%
must hold.

But we already know that
\[
A_{L}\alpha _{L}\left( I\right) L/I=A_{H}\alpha _{H}\left( I\right) H/\left(
1-I\right) ,
\]%
i.e.%
\begin{equation}
\frac{A_{L}}{A_{H}}=\frac{\alpha _{H}\left( I\right) }{\alpha _{L}\left(
I\right) }\frac{H/\left( 1-I\right) }{L/I}.  \label{c_2}
\end{equation}%
It is easy to see that
\[
\frac{\alpha _{H}\left( I\right) }{\alpha _{L}\left( I\right) }\geq \frac{%
\alpha _{H}\left( \tilde{I}_{2}\right) }{\alpha _{L}\left( \tilde{I}%
_{2}\right) },\text{ and }\frac{H/\left( 1-I\right) }{L/I}>\frac{H/\left( 1-%
\tilde{I}_{2}\right) }{L/\left( \tilde{I}_{2}-\tilde{I}_{1}\right) }.
\]%
Then equations (\ref{c_1}) and (\ref{c_2}) can not hold at the same time. By
contradiction, we know that $\tilde{I}_{2}>I$. Wage premium is determined by
the productivity difference at the threshold%
\[
\frac{\tilde{W}_{H}}{\tilde{W}_{L}}=\frac{A_{H}\alpha _{H}\left( \tilde{I}%
_{2}\right) }{A_{L}\alpha _{L}\left( \tilde{I}_{2}\right) }>\frac{%
A_{H}\alpha _{H}\left( I\right) }{A_{L}\alpha _{L}\left( I\right) }.
\]%
In sum, we have proved that a shift of threshold and an increase in wage
premium.

\subsection{Proof of the Model with Deep Learning AI and LLMs AI Technology}

Now we know the task specific productivity profiles are%
\[
\alpha _{L}\left( i\right) =\left( 1-i\right) ^{\eta },
\]%
\[
\alpha _{H}\left( i\right) =i^{\eta },
\]%
\[
\alpha _{M}\left( i\right) =\left( 1-C-\frac{a}{N\left( i\right) ^{\alpha }}-%
\frac{b}{G^{\beta }}\right) ^{\eta }.
\]

The competitive equilibrium can be characterized by the following conditions:

\begin{enumerate}
\item Law of one price for skills%
\[
\hat{W}_{L}=\hat{p}\left( i\right) A_{L}\alpha _{L}\left( i\right) \text{
for any }i<\hat{I}_{1},
\]%
\[
\ F=\hat{p}\left( i\right) A_{M}\alpha _{M}\left( i\right) \text{\ for any }%
\hat{I}_{3}>i>\hat{I}_{2},
\]%
\[
\hat{W}_{H}\hat{=}\tilde{p}\left( i\right) A_{H}\alpha _{H}\left( i\right) \
\text{\ for any }i>\hat{I}_{3}\text{ and }\hat{I}_{2}>i>\hat{I}_{1}.
\]

\item Goods market (or task market) clearing condition demands:%
\[
\hat{p}\left( i\right) \hat{y}\left( i\right) =\hat{p}\left( i^{\prime
}\right) \tilde{y}\left( i^{\prime }\right) =\hat{Y}.
\]

In the four intervals, the following equations are satisfied%
\[
\hat{p}\left( i\right) A_{L}\alpha _{L}\left( i\right) \hat{l}\left(
i\right) =\hat{p}\left( i^{\prime }\right) A_{L}\alpha _{L}\left( i^{\prime
}\right) \hat{l}\left( i^{\prime }\right) ,
\]%
\[
\hat{p}\left( i\right) A_{H}\alpha _{H}\left( i\right) \hat{h}\left(
i\right) =\hat{p}\left( i^{\prime }\right) A_{H}\alpha _{H}\left( i^{\prime
}\right) \hat{h}\left( i^{\prime }\right) ,
\]%
\[
\hat{p}\left( i\right) A_{M}\alpha _{M}\left( i\right) \hat{m}\left(
i\right) =\hat{p}\left( i^{\prime }\right) A_{M}\alpha _{M}\left( i^{\prime
}\right) \hat{m}\left( i^{\prime }\right) .
\]

Combining with the condition of the law of one price for skills, we obtain%
\[
\hat{m}\left( i\right) =\hat{m}\left( i^{\prime }\right) \text{ for any }%
\hat{I}_{3}>i>\hat{I}_{2},
\]%
\[
\hat{l}\left( i\right) =\hat{l}\left( i^{\prime }\right) \text{ for any }i<%
\hat{I}_{1},
\]

similarly%
\[
\hat{h}\left( i\right) =\hat{h}\left( i^{\prime }\right) \text{ for any }i>%
\hat{I}_{3}\text{ and }\hat{I}_{2}>i>\hat{I}_{1}
\]

\item No arbitrage condition across skills and models:
\[
A_{L}\alpha _{L}\left( \hat{I}_{1}\right) \hat{l}\left( \hat{I}_{1}\right)
=A_{H}\alpha _{H}\left( \hat{I}_{1}\right) \hat{h}\left( \hat{I}_{1}\right)
,
\]%
\[
A_{L}\alpha _{L}\left( \hat{I}_{1}\right) \hat{l}\left( \hat{I}_{1}\right)
=A_{M}\alpha _{M}\left( \hat{I}_{1}\right) \hat{m}\left( \hat{I}_{1}\right)
,
\]

\item Labor market clearing condition:%
\[
\int_{0}^{I}\hat{l}\left( i\right) di=L,
\]%
\[
\int_{I}^{1}\hat{h}\left( i\right) di=H.
\]
\end{enumerate}

With previous conditions, we obtain that
\[
\hat{l}\left( i\right) =L/\hat{I}_{1},
\]%
\[
\hat{h}\left( i\right) =H/\left( 1-\hat{I}_{3}+\hat{I}_{2}-\hat{I}%
_{1}\right) .
\]

We can prove the proposition 3 by contradiction. If $\hat{I}_{1}\geq I$, we
know that
\[
H/\left( 1-\hat{I}_{3}+\hat{I}_{2}-\hat{I}_{1}\right) >H/\left( 1-I\right) ,
\]%
and%
\[
L/\hat{I}_{1}\leq L/I.
\]%
With the No arbitrage condition at the threshold, we know that
\[
A_{L}\alpha _{L}\left( \hat{I}_{1}\right) \hat{l}\left( \hat{I}_{1}\right)
=A_{H}\alpha _{H}\left( \hat{I}_{1}\right) \hat{h}\left( \hat{I}_{1}\right)
,
\]%
i.e.%
\begin{equation}
\frac{A_{L}}{A_{H}}=\frac{\alpha _{H}\left( \hat{I}_{1}\right) }{\alpha
_{L}\left( \hat{I}_{1}\right) }\frac{H/\left( 1-\hat{I}_{3}+\hat{I}_{2}-\hat{%
I}_{1}\right) }{L/\hat{I}_{1}}  \label{c_3}
\end{equation}%
must hold.

But we already know that
\[
A_{L}\alpha _{L}\left( I\right) L/I=A_{H}\alpha _{H}\left( I\right) H/\left(
1-I\right) ,
\]%
i.e.%
\begin{equation}
\frac{A_{L}}{A_{H}}=\frac{\alpha _{H}\left( I\right) }{\alpha _{L}\left(
I\right) }\frac{H/\left( 1-I\right) }{L/I}.  \label{c_4}
\end{equation}%
It is easy to see that
\[
\frac{\alpha _{H}\left( I\right) }{\alpha _{L}\left( I\right) }\leq \frac{%
\alpha _{H}\left( \hat{I}_{1}\right) }{\alpha _{L}\left( \hat{I}_{1}\right) }%
,\text{ and }\frac{H/\left( 1-I\right) }{L/I}<\frac{H/\left( 1-\hat{I}_{3}+%
\hat{I}_{2}-\hat{I}_{1}\right) }{L/\hat{I}_{1}}.
\]%
Then equations (\ref{c_3}) and (\ref{c_4}) can not hold at the same time. By
contradiction, we know that $\hat{I}_{1}<I$. Wage premium is determined by
the productivity difference at the threshold%
\[
\frac{\hat{W}_{H}}{\hat{W}_{L}}=\frac{A_{H}\alpha _{H}\left( \hat{I}%
_{1}\right) }{A_{L}\alpha _{L}\left( \hat{I}_{1}\right) }<\frac{A_{H}\alpha
_{H}\left( I\right) }{A_{L}\alpha _{L}\left( I\right) }.
\]%
In sum, we have proved that a shift of threshold and a decrease in wage
premium.

\renewcommand{\thesection}{Appendix B}
\section{Methodology Notes}\label{appendix_b}

\renewcommand{\thesection}{B}



\subsection{LLMs Scoring Details} 
\ 
\newline
The scores fall into the following categories: 
\begin{itemize}
\item E0 indicates no exposure as the LLMs are neither sufficiently useful for the occupation, nor considered a result of the intrinsic nature of the occupation, (e.g., involving physical activities);
\item E1 is applied if a 50 reduction in completion time is already feasible with the existing large language model interfaces; 
\item E2 is applied if such a productivity gain is feasible, but only once the current capabilities of the model can be deployed through applications with further inputs or if it is trained on domain-specific issues or data; 
\item E3 is applied when the productivity increase would require image processing capabilities in addition to current text processing.
\end{itemize}

\subsection{Expert Rubrics} 
\ 
\newline
You are an expert evaluator of the potential for large language models to replace human labor. Large language models are deep learning models used to process and generate natural language text. The latest large language models can generate and describe images and videos based on natural language texts. In this context, you are asked to score each occupation based on whether the occupation could reduce the need for human labor time and participation to achieve the same output or effect in the same amount of time with the aid of large language models.

The scores range from 0 to 1:
\begin{itemize}
\item 0 means the occupation cannot reduce human labor input with the help of large language models.
\item 0.2 means human labor input could be reduced by 20\%.
\item 0.4 means human labor input could be reduced by 40\%.
\item 0.6 means human labor input could be reduced by 60\%.
\item 0.8 means human labor input could be reduced by 80\%.
\item 1 means human labor input could be reduced by 100\%, meaning the occupation would no longer require human participation.
\end{itemize}

Your scores represent the potential proportion of human labor input that could be saved by large language models for each occupation. Please score carefully based on the current capabilities of large language models and what you think is possible in the future.

\renewcommand{\thesection}{Appendix C}
\section{Appendix Figures}\label{appendix_c}

\renewcommand{\thesection}{C}

\begin{figure}[htb]
\begin{centering}
\includegraphics[width=0.6\linewidth]{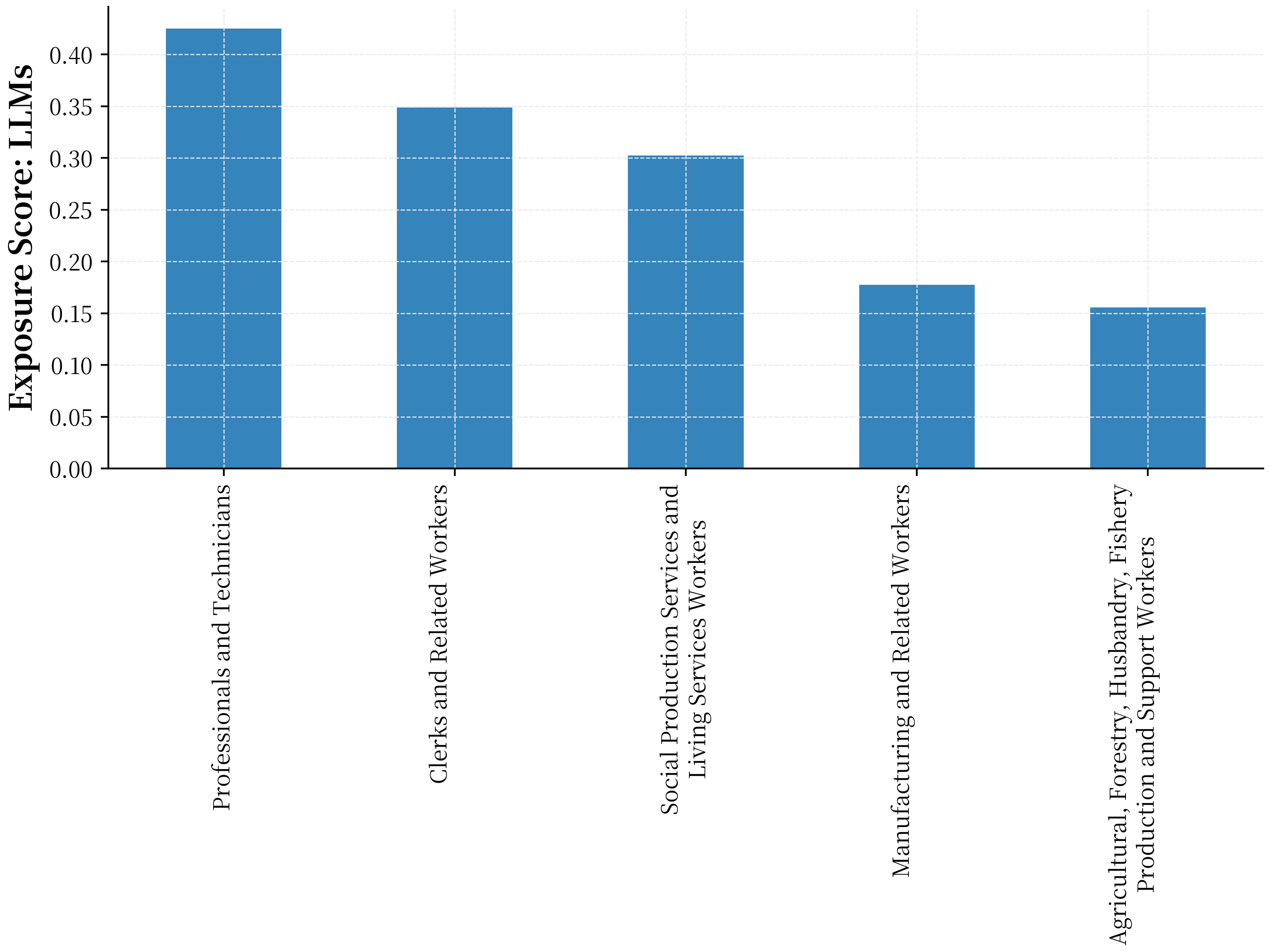}
\par\end{centering}
\caption{Exposure Score on Large-category Level: LLMs.}
\label{fig:large_score_GLM2+Internlm+GPT}
\vspace{-1.2em}
\end{figure}
\begin{figure}[htb]
\begin{centering}
\includegraphics[width=0.6\linewidth]{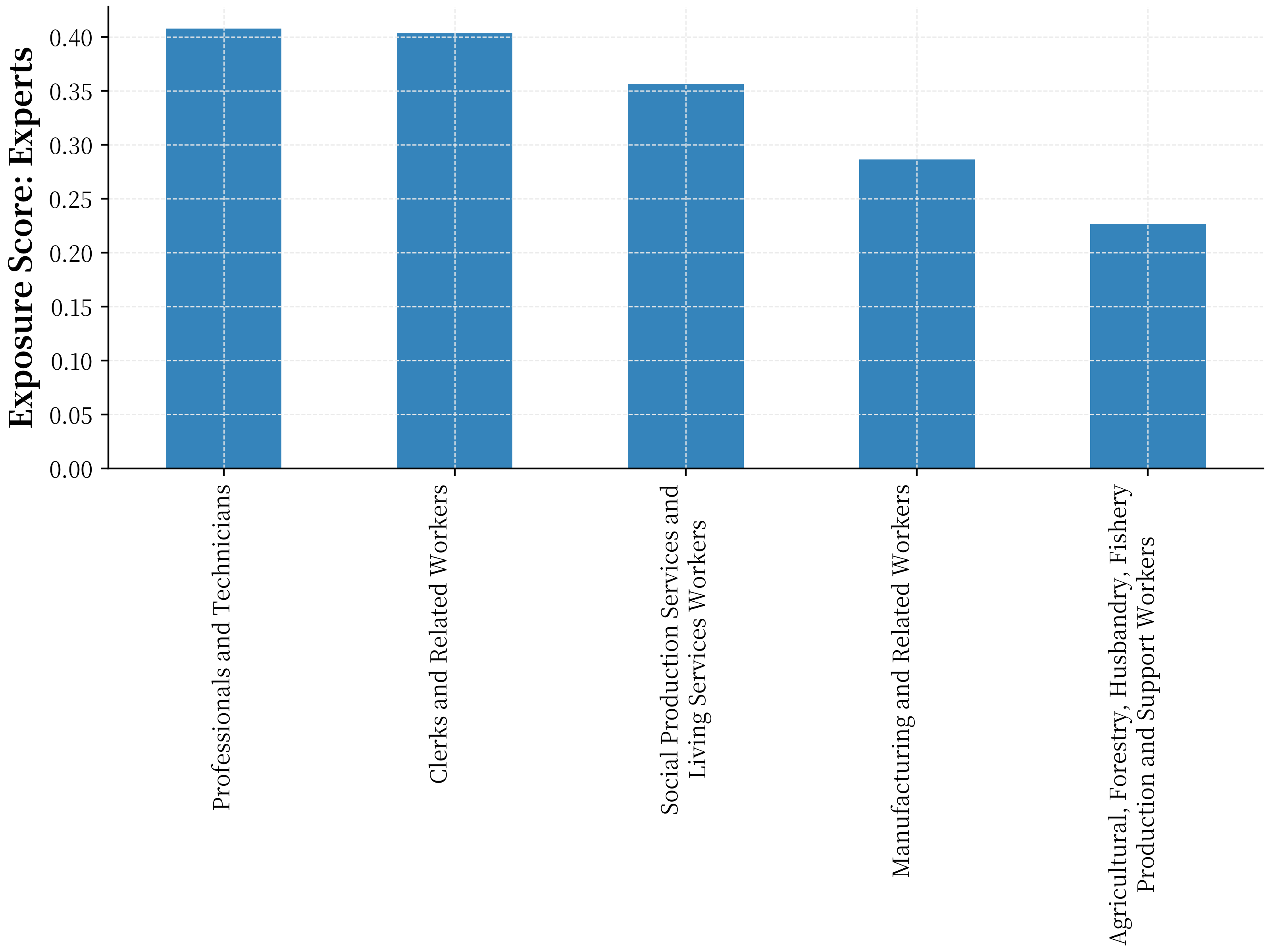}
\par\end{centering}
\caption{Exposure Score on Large-category Level: Experts.}
\label{fig:large_score_expert}
\vspace{-1.2em}
\end{figure}
\begin{figure*}[htb]
\begin{centering}
\includegraphics[width=0.95\linewidth]{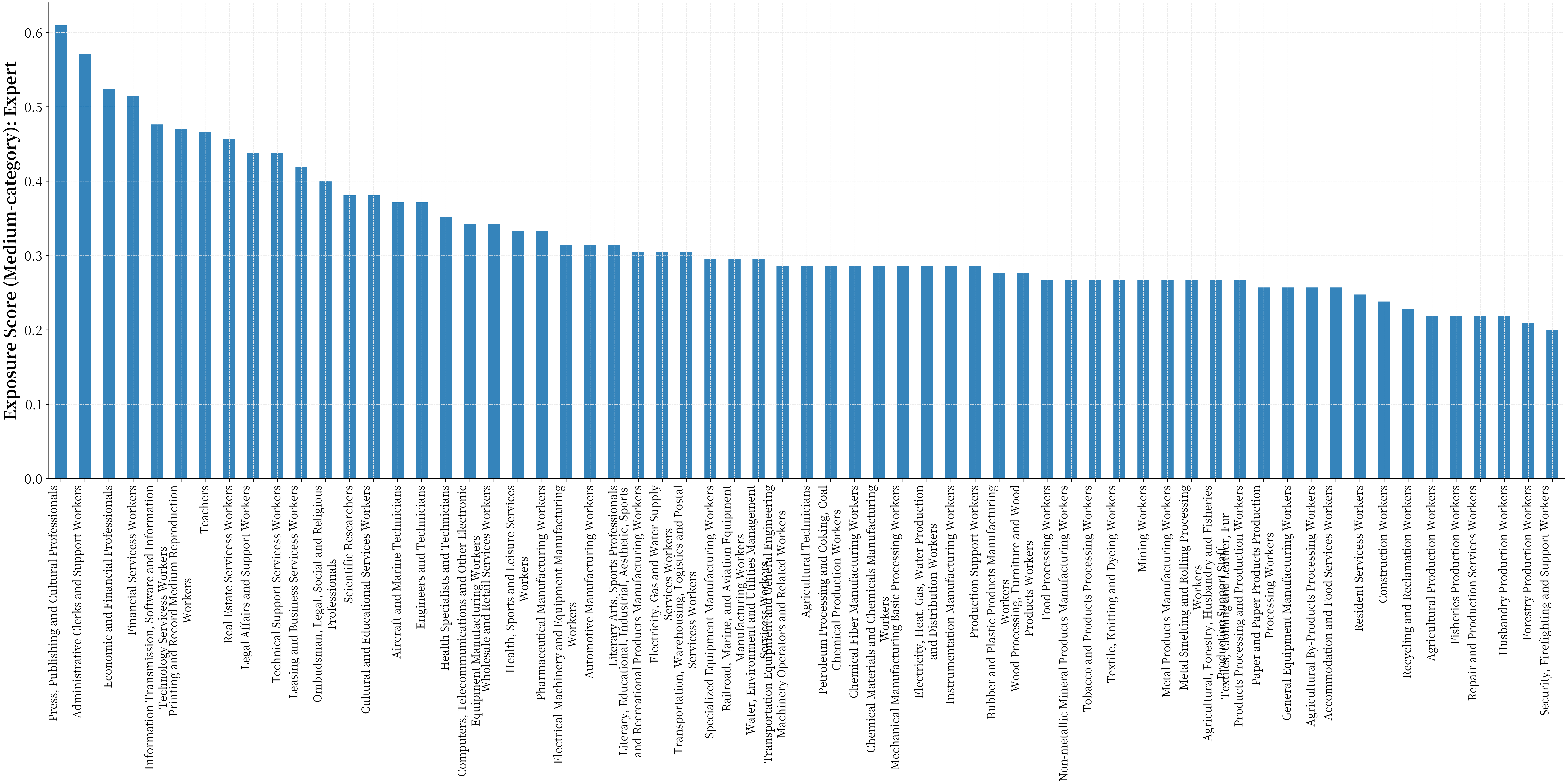}
\par\end{centering}
\caption{AI Medium-category Occupation Exposure: Expert.}
\label{fig:medium_score_expert}
\end{figure*}
\begin{figure}[htb]
\begin{centering}
\includegraphics[width=0.6\linewidth]{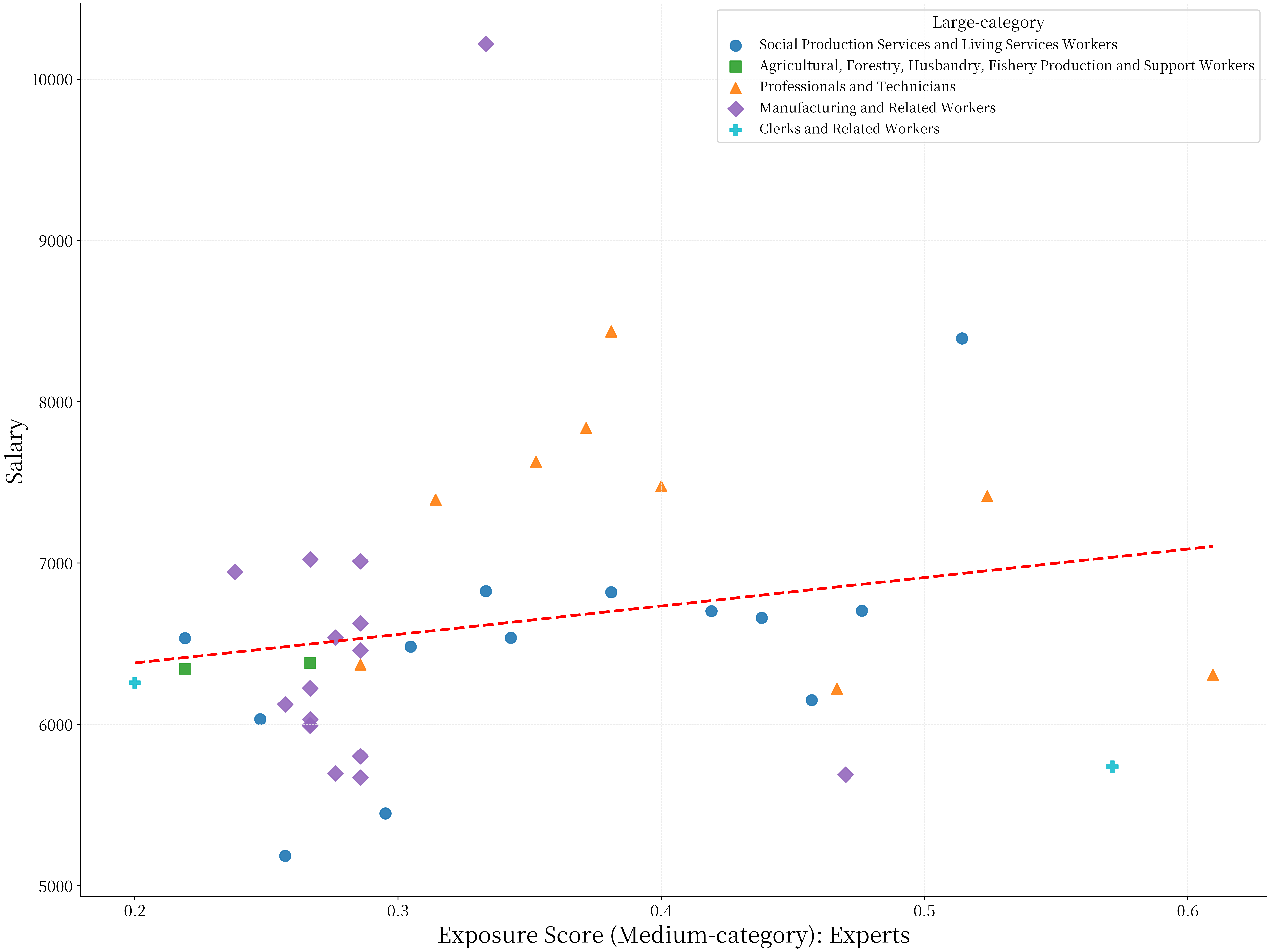}
\par\end{centering}
\caption{Salary and Exposure Score (Medium-category): Experts.}
\label{fig:salary_Expert}
\end{figure}
\begin{figure}[htb]
\begin{centering}
\includegraphics[width=0.6\linewidth]{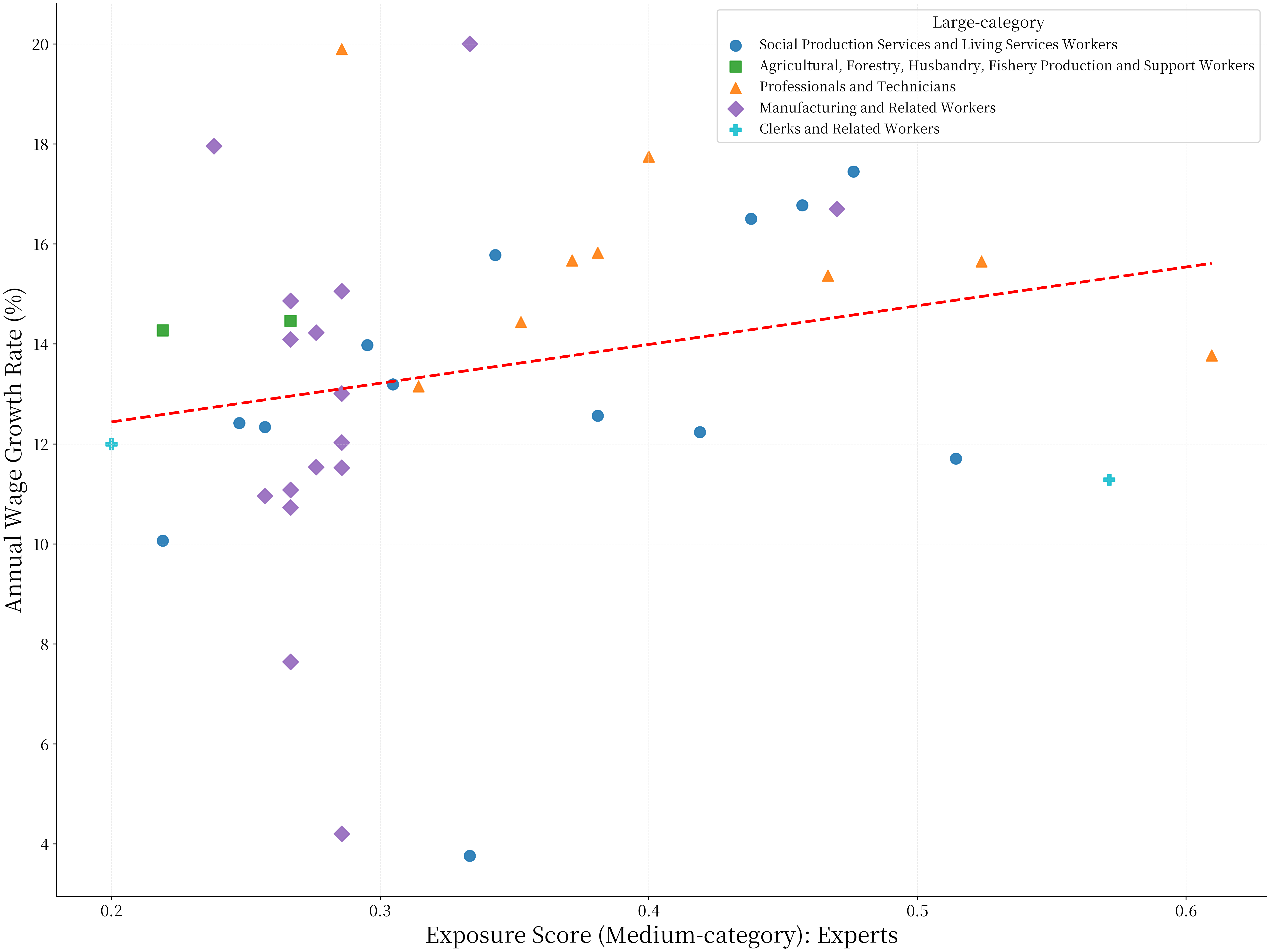}
\par\end{centering}
\caption{Annual Wage Growth Rate and Exposure Score(Medium-category): Experts.}
\label{fig:wage_growth_Expert}
\end{figure}
\begin{figure*}[htb]
\begin{centering}
\includegraphics[width=0.9\linewidth]{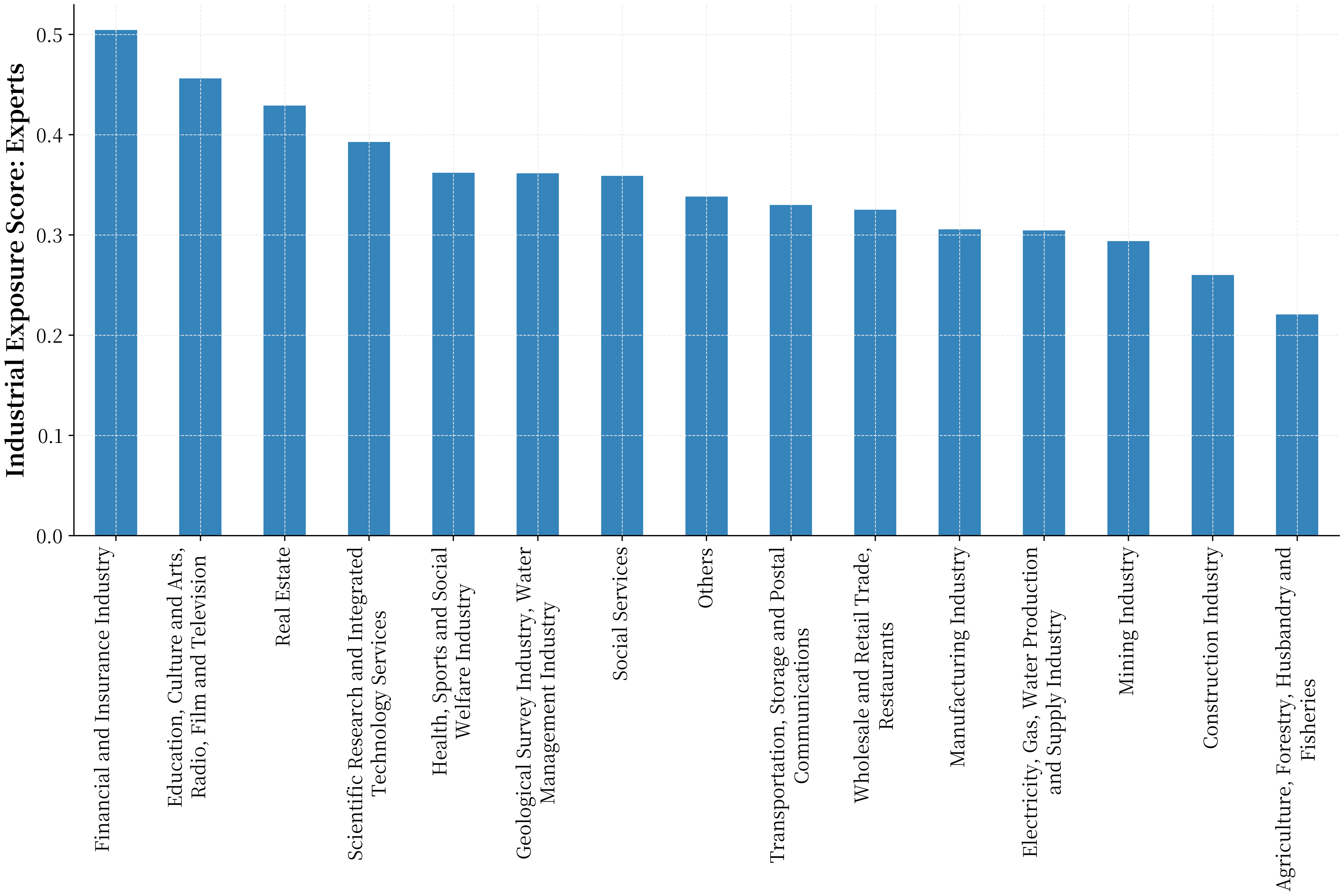}
\par\end{centering}
\caption{Industrial Exposure Score: Experts.}
\label{fig: ind_score_Expert}
\vspace{-1.2em}
\end{figure*}

\begin{figure}[htb]
\begin{centering}
\includegraphics[width=0.5\linewidth]{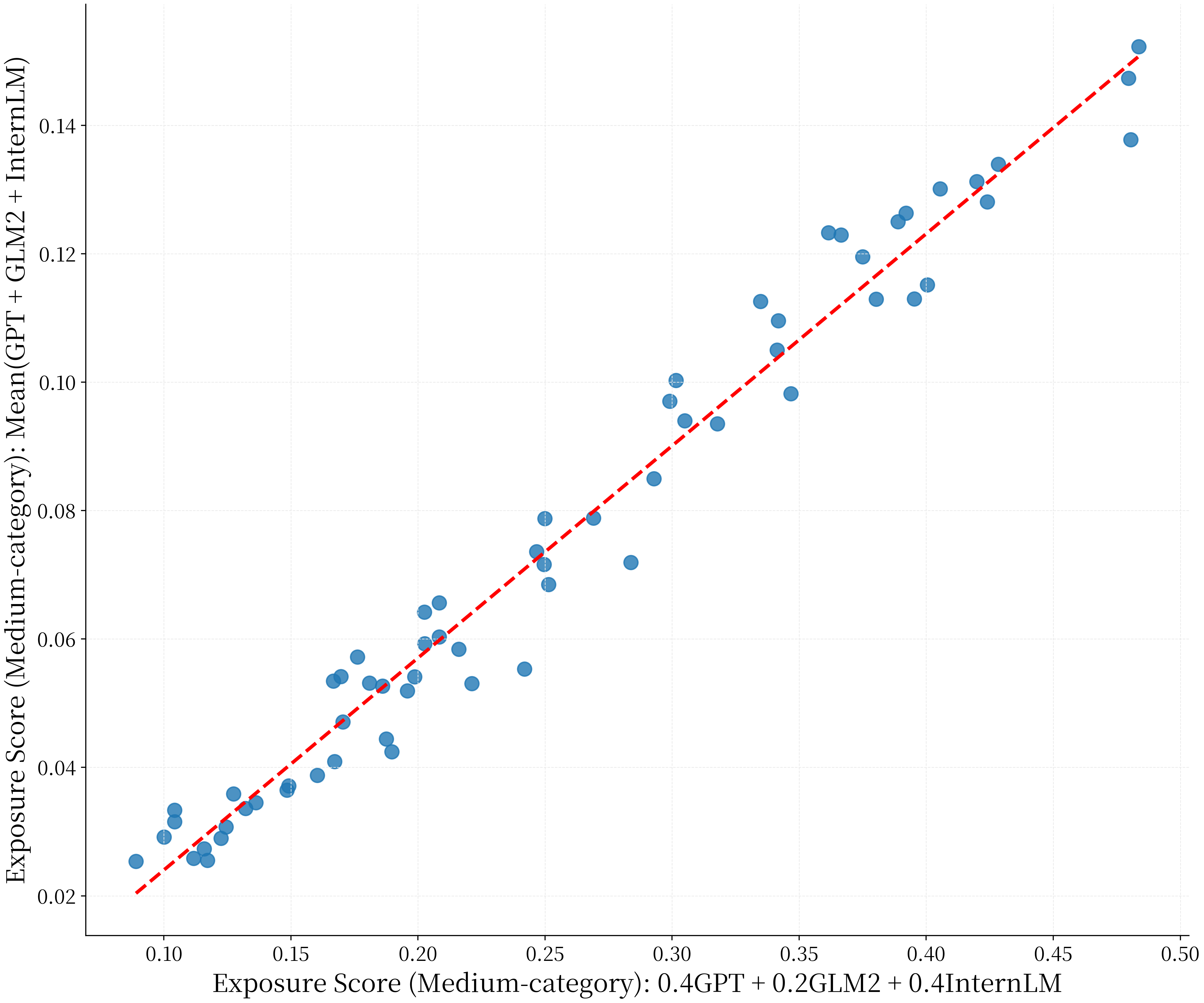}
\par\end{centering}
\caption{Exposure Scores with Different Weighted Assignments}
\label{fig:cor_GLM2+Internlm+GPT_Expert}
\vspace{-1.2em}
\end{figure}
\begin{figure}[htb]
\begin{centering}
\includegraphics[width=0.65\linewidth]{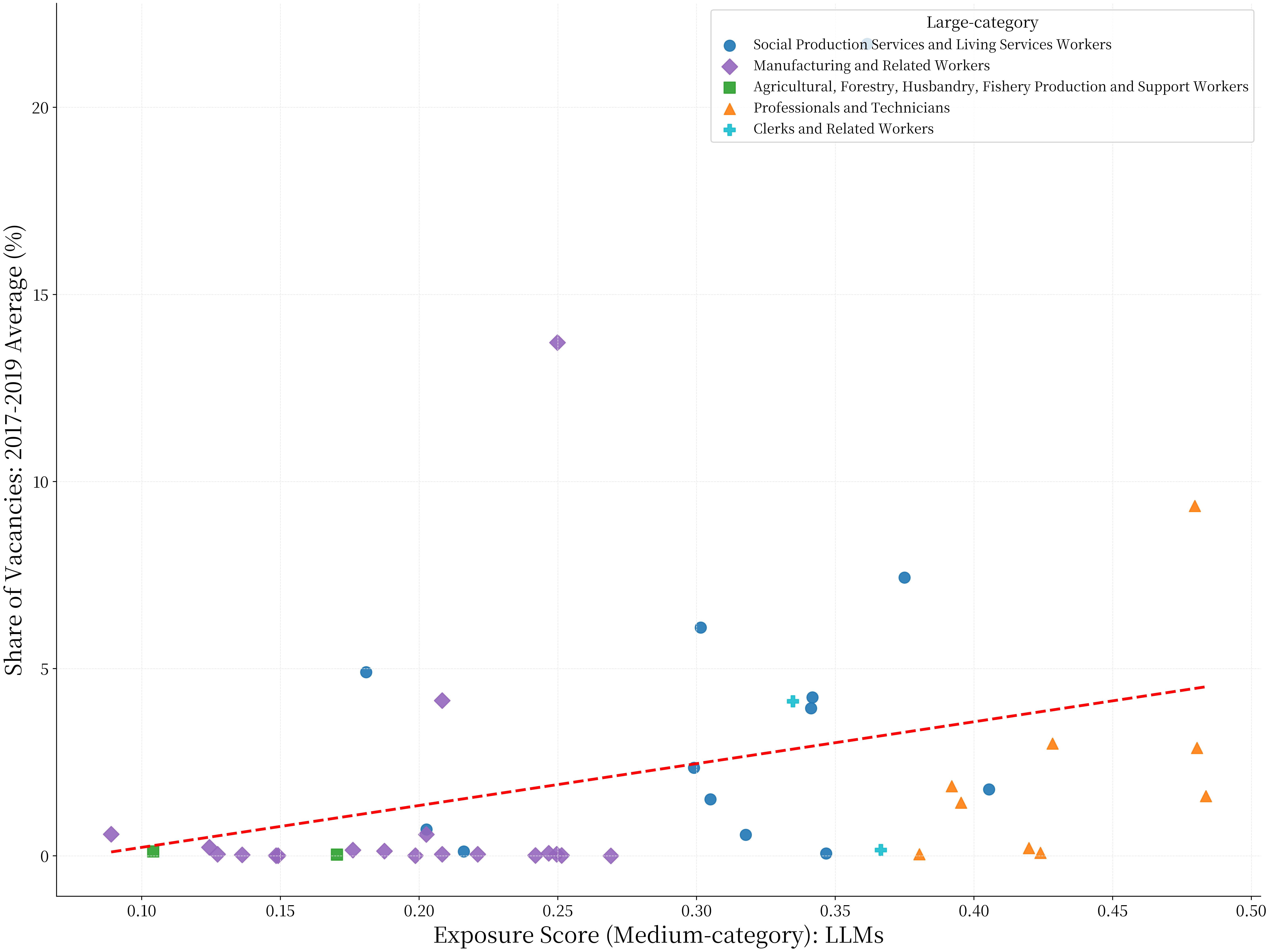}
\par\end{centering}
\caption{Share of Vacancies and Exposure Score(2017-2019): LLMs.}
\label{fig: share of vacancy llms}
\end{figure}
\begin{figure}[htb]
\begin{centering}
\includegraphics[width=0.65\linewidth]{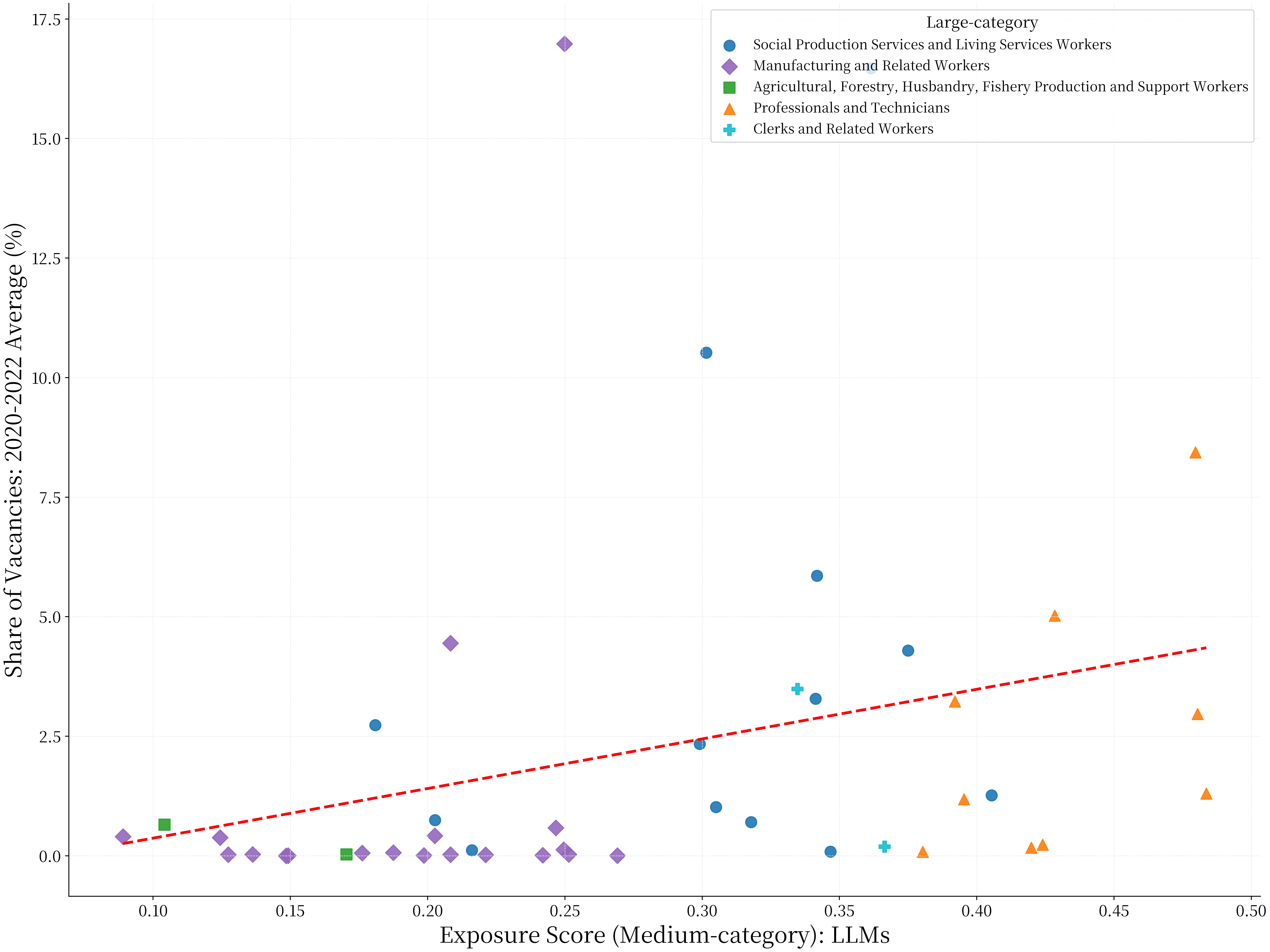}
\par\end{centering}
\caption{Share of Vacancies and Exposure Score(2020-2022): LLMs.}
\label{fig: share of vacancy llms}
\end{figure}

\onecolumn

\renewcommand{\thesection}{Appendix D}
\section{Appendix Tables}\label{appendix_d}

\renewcommand{\thesection}{D}

\begin{center}
\begin{longtable}
{|p{3.94em}|p{13.315em}|l|l|l|l|l|}
    \caption{Full List if Medium-category Occupations of AI Exposure Score}
    \label{tab:table medium score} \\
    \toprule
    \textbf{Occupation Code} & \textbf{Medium-category Occupation Title} & \multicolumn{1}{p{3.125em}|}{\textbf{Expert}} & \multicolumn{1}{p{3.125em}|}{\textbf{GLM}} & \multicolumn{1}{p{3.125em}|}{\textbf{GPT4}} & \multicolumn{1}{p{4.065em}|}{\textbf{InternLM}} & \multicolumn{1}{p{5.125em}|}{\textbf{GLM + InternLM + GPT4}} \\
    \midrule
    2-01  & Scientific Researchers & 0.3810 & 0.6227 & 0.3472 & 0.3021 & 0.4240 \\
    \midrule
    2-02  & Engineers and Technicians & 0.3714 & 0.5612 & 0.3600 & 0.3640 & 0.4284 \\
    \midrule
    2-03  & Agricultural Technicians & 0.2857 & 0.5881 & 0.2308 & 0.3221 & 0.3803 \\
    \midrule
    2-04  & Aircraft and Marine Technicians & 0.3714 & 0.6235 & 0.0893 & 0.1384 & 0.2837 \\
    \midrule
    2-05  & Health Specialists and Technicians & 0.3524 & 0.6776 & 0.2778 & 0.2307 & 0.3954 \\
    \midrule
    2-06  & Economic and Financial Professionals & 0.5238 & 0.6673 & 0.4467 & 0.3248 & 0.4796 \\
    \midrule
    2-07  & Ombudsman, Legal, Social and Religious Professionals & 0.4000 & 0.5505 & 0.4808 & 0.2284 & 0.4199 \\
    \midrule
    2-08  & Teachers & 0.4667 & 0.8164 & 0.5313 & 0.0938 & 0.4805 \\
    \midrule
    2-09  & Literary Arts, Sports Professionals & 0.3143 & 0.4573 & 0.5196 & 0.1991 & 0.3920 \\
    \midrule
    2-10  & Press, Publishing and Cultural Professionals & 0.6095 & 0.6176 & 0.5179 & 0.3155 & 0.4836 \\
    \midrule
    3-01  & Administrative Clerks and Support Workers & 0.5714 & 0.3196 & 0.5568 & 0.1278 & 0.3348 \\
    \midrule
    3-02  & Security, Firefighting and Support Workers & 0.2000 & 0.3551 & 0.4545 & 0.2898 & 0.3665 \\
    \midrule
    3-03  & Legal Affairs and Support Workers & 0.4381 & 0.4583 & 0.5000 & 0.2083 & 0.3889 \\
    \midrule
    4-01  & Wholesale and Retail Services Workers & 0.3429 & 0.3199 & 0.5515 & 0.2132 & 0.3615 \\
    \midrule
    4-02  & Transportation, Warehousing, Logistics and Postal Servicess Workers & 0.3048 & 0.3047 & 0.5160 & 0.0838 & 0.3015 \\
    \midrule
    4-03  & Accommodation and Food Services Workers & 0.2571 & 0.4570 & 0.4531 & 0.2148 & 0.3750 \\
    \midrule
    4-04  & Information Transmission, Software and Information Technology Servicess Workers & 0.4762 & 0.4074 & 0.3241 & 0.2940 & 0.3418 \\
    \midrule
    4-05  & Financial Servicess Workers & 0.5143 & 0.4808 & 0.4327 & 0.3029 & 0.4054 \\
    \midrule
    4-06  & Real Estate Servicess Workers & 0.4571 & 0.3393 & 0.3929 & 0.1652 & 0.2991 \\
    \midrule
    4-07  & Leasing and Business Servicess Workers & 0.4190 & 0.4724 & 0.3669 & 0.1845 & 0.3413 \\
    \midrule
    4-08  & Technical Support Servicess Workers & 0.4381 & 0.4199 & 0.3066 & 0.1885 & 0.3050 \\
    \midrule
    4-09  & Water, Environment and Utilities Management Servicess Workers & 0.2952 & 0.3267 & 0.1583 & 0.1229 & 0.2027 \\
    \midrule
    4-10  & Resident Servicess Workers & 0.2476 & 0.2885 & 0.1417 & 0.1125 & 0.1809 \\
    \midrule
    4-11  & Electricity, Gas and Water Supply Services Workers & 0.3048 & 0.6753 & 0.1458 & 0.3802 & 0.4005 \\
    \midrule
    4-12  & Repair and Production Services Workers & 0.2190 & 0.4203 & 0.1250 & 0.1029 & 0.2161 \\
    \midrule
    4-13  & Cultural and Educational Services Workers & 0.3810 & 0.5036 & 0.1685 & 0.2813 & 0.3178 \\
    \midrule
    4-14  & Health, Sports and Leisure Services Workers & 0.3333 & 0.6070 & 0.1595 & 0.2737 & 0.3467 \\
    \midrule
    5-01  & Agricultural Production Workers & 0.2190 & 0.3259 & 0.1964 & 0.0357 & 0.1860 \\
    \midrule
    5-02  & Forestry Production Workers & 0.2095 & 0.2054 & 0.2321 & 0.0714 & 0.1696 \\
    \midrule
    5-03  & Husbandry Production Workers & 0.2190 & 0.1518 & 0.1429 & 0.0179 & 0.1042 \\
    \midrule
    5-04  & Fisheries Production Workers & 0.2190 & 0.2852 & 0.0000 & 0.0625 & 0.1159 \\
    \midrule
    5-05  & Agricultural, Forestry, Husbandry and Fisheries Production Support Staff & 0.2667 & 0.3163 & 0.0650 & 0.1300 & 0.1704 \\
    \midrule
    6-01  & Agricultural By-Products Processing Workers & 0.2571 & 0.3958 & 0.1750 & 0.0167 & 0.1958 \\
    \midrule
    6-02  & Food Processing Workers & 0.2667 & 0.4807 & 0.2054 & 0.0681 & 0.2514 \\
    \midrule
    6-03  & Tobacco and Products Processing Workers & 0.2667 & 0.4826 & 0.3333 & 0.0625 & 0.2928 \\
    \midrule
    6-04  & Textile, Knitting and Dyeing Workers & 0.2667 & 0.2259 & 0.1193 & 0.0369 & 0.1274 \\
    \midrule
    6-05  & Textiles, Clothing and Leather, Fur Products Processing and Production Workers & 0.2667 & 0.1534 & 0.0682 & 0.0455 & 0.0890 \\
    \midrule
    6-06  & Wood Processing, Furniture and Wood Products Workers & 0.2762 & 0.1989 & 0.2727 & 0.0568 & 0.1761 \\
    \midrule
    6-07  & Paper and Paper Products Production Processing Workers & 0.2571 & 0.1979 & 0.2292 & 0.0729 & 0.1667 \\
    \midrule
    6-08  & Printing and Record Medium Reproduction Workers & 0.4700 & 0.2656 & 0.2813 & 0.0781 & 0.2083 \\
    \midrule
    6-09  & Literary, Educational, Industrial, Aesthetic, Sports and Recreational Products Manufacturing Workers  & 0.3048 & 0.2997 & 0.0481 & 0.0609 & 0.1362 \\
    \midrule
    6-10  & Petroleum Processing and Coking, Coal Chemical Production Workers & 0.2857 & 0.5016 & 0.0000 & 0.0674 & 0.1897 \\
    \midrule
    6-11  & Chemical Materials and Chemicals Manufacturing Workers & 0.2857 & 0.5307 & 0.0000 & 0.1327 & 0.2211 \\
    \midrule
    6-12  & Pharmaceutical Manufacturing Workers & 0.3333 & 0.6215 & 0.0000 & 0.1042 & 0.2419 \\
    \midrule
    6-13  & Chemical Fiber Manufacturing Workers & 0.2857 & 0.3203 & 0.0000 & 0.0313 & 0.1172 \\
    \midrule
    6-14  & Rubber and Plastic Products Manufacturing Workers & 0.2762 & 0.4583 & 0.0000 & 0.1042 & 0.1875 \\
    \midrule
    6-15  & Non-metallic Mineral Products Manufacturing Workers & 0.2667 & 0.2829 & 0.0174 & 0.0349 & 0.1118 \\
    \midrule
    6-16  & Mining Workers & 0.2667 & 0.2885 & 0.0000 & 0.1080 & 0.1322 \\
    \midrule
    6-17  & Metal Smelting and Rolling Processing Workers & 0.2667 & 0.3379 & 0.0660 & 0.0436 & 0.1492 \\
    \midrule
    6-18  & Mechanical Manufacturing Basic Processing Workers & 0.2857 & 0.3448 & 0.2500 & 0.0302 & 0.2083 \\
    \midrule
    6-19  & Metal Products Manufacturing Workers & 0.2667 & 0.3438 & 0.0625 & 0.0391 & 0.1484 \\
    \midrule
    6-20  & General Equipment Manufacturing Workers & 0.2571 & 0.2856 & 0.0000 & 0.0877 & 0.1244 \\
    \midrule
    6-21  & Specialized Equipment Manufacturing Workers & 0.2952 & 0.3006 & 0.0000 & 0.0670 & 0.1225 \\
    \midrule
    6-22  & Automotive Manufacturing Workers & 0.3143 & 0.3813 & 0.0000 & 0.1000 & 0.1604 \\
    \midrule
    6-23  & Railroad, Marine, and Aviation Equipment Manufacturing Workers & 0.2952 & 0.1629 & 0.0000 & 0.1373 & 0.1001 \\
    \midrule
    6-24  & Electrical Machinery and Equipment Manufacturing Workers & 0.3143 & 0.3898 & 0.0197 & 0.0921 & 0.1672 \\
    \midrule
    6-25  & Computers, Telecommunications and Other Electronic Equipment Manufacturing Workers & 0.3429 & 0.3801 & 0.1212 & 0.0947 & 0.1987 \\
    \midrule
    6-26  & Instrumentation Manufacturing Workers & 0.2857 & 0.4323 & 0.1250 & 0.2500 & 0.2691 \\
    \midrule
    6-27  & Recycling and Reclamation Workers & 0.2286 & 0.1250 & 0.1250 & 0.0625 & 0.1042 \\
    \midrule
    6-28  & Electricity, Heat, Gas, Water Production and Distribution Workers & 0.2857 & 0.4236 & 0.2407 & 0.0845 & 0.2496 \\
    \midrule
    6-29  & Construction Workers & 0.2381 & 0.2525 & 0.2072 & 0.1480 & 0.2026 \\
    \midrule
    6-30  & Transportation Equipment and General Engineering Machinery Operators and Related Workers & 0.2857 & 0.3760 & 0.2438 & 0.1203 & 0.2467 \\
    \midrule
    6-31  & Production Support Workers & 0.2857 & 0.3179 & 0.2315 & 0.2002 & 0.2499 \\
    \bottomrule
\end{longtable}%
\end{center}

\begin{center}
\begin{longtable}{|p{25.44em}|c|}
    \caption{List of Industry ID}
    \label{tab:table ind id} \\
    \toprule
    \textbf{Ind.} & \multicolumn{1}{p{1.44em}|}{\textbf{ID}} \\
    \midrule
    Education, Culture and Arts, Radio, Film and Television & 1 \\
    \midrule
    Scientific Research and Integrated Technology Services & 2 \\
    \midrule
    Wholesale and Retail Trade, Restaurants & 3 \\
    \midrule
    Social Services & 4 \\
    \midrule
    Others & 5 \\
    \midrule
    Financial and Insurance Industry & 6 \\
    \midrule
    Transportation, Storage and Postal Communications & 7 \\
    \midrule
    Real Estate & 8 \\
    \midrule
    Construction Industry & 9 \\
    \midrule
    Electricity, Gas, Water Production and Supply Industry & 10 \\
    \midrule
    Manufacturing Industry & 11 \\
    \midrule
    Health, Sports and Social Welfare Industry & 12 \\
    \midrule
    Agriculture, Forestry, Husbandry and Fisheries & 13 \\
    \midrule
    Mining Industry & 14 \\
    \midrule
    Geological Survey Industry, Water Management Industry & 15 \\
    \bottomrule
\end{longtable}%
\end{center}

%
\begin{center}
\begin{longtable}
{|p{9em}|p{1em}|p{1em}|p{1em}|p{1em}|p{1em}|p{1em}|p{1em}|p{1em}|p{1em}|p{1em}|p{1em}|p{1em}|p{1em}|p{1em}|p{1em}|}
    \caption{Full List of Occupation Share of Industry}
    \label{tab:table ind occ share} \\
    \toprule
                           Ind.                                                                    \newline{}Occ. & \multicolumn{1}{l|}{\textbf{1}} & \multicolumn{1}{l|}{\textbf{2}} & \multicolumn{1}{l|}{\textbf{3}} & \multicolumn{1}{l|}{\textbf{4}} & \multicolumn{1}{l|}{\textbf{5}} & \multicolumn{1}{l|}{\textbf{6}} & \multicolumn{1}{l|}{\textbf{7}} & \multicolumn{1}{l|}{\textbf{8}} & \multicolumn{1}{l|}{\textbf{9}} & \multicolumn{1}{l|}{\textbf{10}} & \multicolumn{1}{l|}{\textbf{11}} & \multicolumn{1}{l|}{\textbf{12}} & \multicolumn{1}{l|}{\textbf{13}} & \multicolumn{1}{l|}{\textbf{14}} & \multicolumn{1}{l|}{\textbf{15}} \\
    \midrule
    Transporta-tion, Warehousing, Logistics and Postal Services Workers & 0\%   & 0\%   & 1\%   & 7\%   & 8\%   & 1\%   & 72\%  & 0\%   & 1\%   & 1\%   & 3\%   & 1\%   & 0\%   & 4\%   & 0\% \\
    \midrule
    Accommoda-tion and Food Services Workers & 0\%   & 0\%   & 21\%  & 10\%  & 5\%   & 1\%   & 0\%   & 1\%   & 0\%   & 1\%   & 1\%   & 0\%   & 0\%   & 0\%   & 5\% \\
    \midrule
    Information Transmission, Software and Information Technology Servicess Workers & 0\%   & 23\%  & 0\%   & 1\%   & 4\%   & 1\%   & 6\%   & 0\%   & 0\%   & 1\%   & 0\%   & 0\%   & 0\%   & 0\%   & 0\% \\
    \midrule
    Repair and Production Services Workers & 0\%   & 7\%   & 0\%   & 2\%   & 5\%   & 0\%   & 2\%   & 0\%   & 0\%   & 2\%   & 1\%   & 0\%   & 0\%   & 1\%   & 0\% \\
    \midrule
    Agricultural Production Workers & 0\%   & 2\%   & 0\%   & 0\%   & 3\%   & 1\%   & 0\%   & 1\%   & 1\%   & 0\%   & 0\%   & 1\%   & 96\%  & 0\%   & 0\% \\
    \midrule
    Printing and Record Medium Reproduction Workers & 0\%   & 0\%   & 0\%   & 0\%   & 0\%   & 0\%   & 0\%   & 0\%   & 0\%   & 0\%   & 1\%   & 0\%   & 0\%   & 0\%   & 0\% \\
    \midrule
    Resident Servicess Workers & 2\%   & 8\%   & 0\%   & 14\%  & 7\%   & 0\%   & 0\%   & 0\%   & 0\%   & 1\%   & 1\%   & 2\%   & 0\%   & 0\%   & 0\% \\
    \midrule
    Engineers and Technicians & 0\%   & 18\%  & 0\%   & 0\%   & 0\%   & 1\%   & 1\%   & 1\%   & 3\%   & 2\%   & 2\%   & 1\%   & 0\%   & 2\%   & 14\% \\
    \midrule
    Construction Workers & 0\%   & 0\%   & 0\%   & 1\%   & 3\%   & 0\%   & 1\%   & 9\%   & 78\%  & 5\%   & 1\%   & 0\%   & 0\%   & 0\%   & 5\% \\
    \midrule
    Wholesale and Retail Services Workers & 1\%   & 3\%   & 69\%  & 14\%  & 17\%  & 3\%   & 4\%   & 9\%   & 3\%   & 1\%   & 3\%   & 8\%   & 0\%   & 2\%   & 5\% \\
    \midrule
    Technical Support Servicess Workers & 3\%   & 2\%   & 0\%   & 0\%   & 2\%   & 0\%   & 0\%   & 1\%   & 0\%   & 0\%   & 1\%   & 0\%   & 0\%   & 0\%   & 18\% \\
    \midrule
    Teachers & 77\%  & 0\%   & 0\%   & 0\%   & 0\%   & 0\%   & 0\%   & 0\%   & 0\%   & 0\%   & 0\%   & 3\%   & 0\%   & 0\%   & 0\% \\
    \midrule
    Cultural and Educational Services Workers & 1\%   & 3\%   & 0\%   & 1\%   & 3\%   & 0\%   & 0\%   & 0\%   & 0\%   & 0\%   & 0\%   & 0\%   & 0\%   & 0\%   & 0\% \\
    \midrule
    Literary Arts, Sports Professionals & 4\%   & 2\%   & 0\%   & 0\%   & 2\%   & 0\%   & 0\%   & 0\%   & 0\%   & 0\%   & 0\%   & 0\%   & 0\%   & 0\%   & 0\% \\
    \midrule
    Press, Publishing and Cultural Professionals & 2\%   & 0\%   & 0\%   & 0\%   & 0\%   & 0\%   & 0\%   & 0\%   & 0\%   & 0\%   & 0\%   & 1\%   & 0\%   & 0\%   & 0\% \\
    \midrule
    Ombudsman, Legal, Social and Religious Professionals & 0\%   & 0\%   & 0\%   & 5\%   & 1\%   & 1\%   & 0\%   & 0\%   & 0\%   & 0\%   & 0\%   & 0\%   & 0\%   & 0\%   & 0\% \\
    \midrule
    Leasing and Business Servicess Workers & 1\%   & 2\%   & 1\%   & 11\%  & 9\%   & 1\%   & 1\%   & 5\%   & 0\%   & 4\%   & 1\%   & 1\%   & 0\%   & 2\%   & 0\% \\
    \midrule
    Economic and Financial Professionals & 2\%   & 12\%  & 2\%   & 4\%   & 5\%   & 26\%  & 2\%   & 11\%  & 1\%   & 1\%   & 3\%   & 2\%   & 0\%   & 1\%   & 5\% \\
    \midrule
    Admini-strative Clerks and Support Workers & 4\%   & 5\%   & 0\%   & 11\%  & 5\%   & 2\%   & 3\%   & 4\%   & 1\%   & 3\%   & 3\%   & 7\%   & 0\%   & 4\%   & 9\% \\
    \midrule
    Health, Sports and Leisure Services Workers & 0\%   & 2\%   & 0\%   & 1\%   & 0\%   & 0\%   & 0\%   & 0\%   & 0\%   & 0\%   & 0\%   & 6\%   & 0\%   & 0\%   & 0\% \\
    \midrule
    Agricultural, Forestry, Husbandry and Fisheries Production Support Staff & 0\%   & 2\%   & 0\%   & 0\%   & 1\%   & 0\%   & 0\%   & 0\%   & 0\%   & 0\%   & 0\%   & 0\%   & 0\%   & 0\%   & 0\% \\
    \midrule
    Security, Firefighting and Support Workers & 0\%   & 2\%   & 0\%   & 1\%   & 0\%   & 0\%   & 0\%   & 0\%   & 0\%   & 0\%   & 0\%   & 0\%   & 0\%   & 0\%   & 0\% \\
    \midrule
    Production Support Workers & 0\%   & 3\%   & 0\%   & 0\%   & 4\%   & 0\%   & 1\%   & 0\%   & 2\%   & 17\%  & 8\%   & 1\%   & 0\%   & 8\%   & 0\% \\
    \midrule
    Scientific Researchers & 0\%   & 3\%   & 0\%   & 0\%   & 0\%   & 1\%   & 0\%   & 0\%   & 0\%   & 0\%   & 0\%   & 0\%   & 0\%   & 0\%   & 0\% \\
    \midrule
    Computers, Telecommunications and Other Electronic Equipment Manufacturing Workers & 0\%   & 2\%   & 0\%   & 0\%   & 2\%   & 0\%   & 0\%   & 0\%   & 0\%   & 1\%   & 6\%   & 0\%   & 0\%   & 0\%   & 0\% \\
    \midrule
    Recycling and Reclamation Workers & 0\%   & 0\%   & 0\%   & 0\%   & 1\%   & 0\%   & 0\%   & 0\%   & 0\%   & 0\%   & 0\%   & 0\%   & 0\%   & 0\%   & 5\% \\
    \midrule
    Agricultural By-Products Processing Workers & 0\%   & 0\%   & 0\%   & 0\%   & 1\%   & 0\%   & 0\%   & 0\%   & 0\%   & 0\%   & 0\%   & 0\%   & 0\%   & 0\%   & 0\% \\
    \midrule
    Literary, Educational, Industrial, Aesthetic, Sports and Recreational Products Manufacturing Workers  & 0\%   & 0\%   & 0\%   & 0\%   & 0\%   & 0\%   & 0\%   & 0\%   & 0\%   & 0\%   & 4\%   & 0\%   & 0\%   & 0\%   & 0\% \\
    \midrule
    Wood Processing, Furniture and Wood Products Workers & 0\%   & 0\%   & 0\%   & 0\%   & 1\%   & 0\%   & 0\%   & 0\%   & 2\%   & 0\%   & 4\%   & 0\%   & 0\%   & 0\%   & 0\% \\
    \midrule
    Mechanical Manufacturing Basic Processing Workers & 0\%   & 0\%   & 0\%   & 0\%   & 1\%   & 0\%   & 0\%   & 0\%   & 1\%   & 1\%   & 6\%   & 0\%   & 0\%   & 0\%   & 0\% \\
    \midrule
    Rubber and Plastic Products Manufacturing Workers & 0\%   & 0\%   & 0\%   & 0\%   & 1\%   & 0\%   & 0\%   & 0\%   & 0\%   & 1\%   & 2\%   & 0\%   & 0\%   & 0\%   & 0\% \\
    \midrule
    Water, Environment and Utilities Management Servicess Workers & 0\%   & 0\%   & 1\%   & 10\%  & 3\%   & 0\%   & 0\%   & 0\%   & 0\%   & 1\%   & 0\%   & 20\%  & 0\%   & 0\%   & 14\% \\
    \midrule
    Textile, Knitting and Dyeing Workers & 0\%   & 0\%   & 0\%   & 0\%   & 1\%   & 0\%   & 0\%   & 0\%   & 0\%   & 1\%   & 4\%   & 0\%   & 0\%   & 0\%   & 0\% \\
    \midrule
    Textiles, Clothing and Leather, Fur Products Processing and Production Workers & 0\%   & 0\%   & 0\%   & 0\%   & 1\%   & 0\%   & 0\%   & 0\%   & 0\%   & 0\%   & 13\%  & 0\%   & 0\%   & 0\%   & 0\% \\
    \midrule
    Metal Products Manufacturing Workers & 0\%   & 0\%   & 0\%   & 0\%   & 0\%   & 0\%   & 0\%   & 0\%   & 0\%   & 0\%   & 7\%   & 0\%   & 0\%   & 0\%   & 0\% \\
    \midrule
    Non-metallic Mineral Products Manufacturing Workers & 0\%   & 0\%   & 0\%   & 0\%   & 1\%   & 0\%   & 0\%   & 0\%   & 2\%   & 0\%   & 6\%   & 0\%   & 0\%   & 0\%   & 0\% \\
    \midrule
    Food Processing Workers & 0\%   & 0\%   & 1\%   & 0\%   & 1\%   & 0\%   & 0\%   & 0\%   & 0\%   & 0\%   & 1\%   & 0\%   & 0\%   & 0\%   & 0\% \\
    \midrule
    Health Specialists and Technicians & 0\%   & 0\%   & 0\%   & 1\%   & 0\%   & 0\%   & 0\%   & 0\%   & 0\%   & 0\%   & 0\%   & 41\%  & 0\%   & 1\%   & 0\% \\
    \midrule
    Real Estate Servicess Workers & 0\%   & 0\%   & 0\%   & 4\%   & 0\%   & 0\%   & 0\%   & 56\%  & 0\%   & 0\%   & 0\%   & 0\%   & 0\%   & 0\%   & 0\% \\
    \midrule
    Forestry Production Workers & 0\%   & 0\%   & 0\%   & 0\%   & 0\%   & 0\%   & 0\%   & 0\%   & 0\%   & 0\%   & 0\%   & 0\%   & 0\%   & 0\%   & 0\% \\
    \midrule
    Electricity, Gas and Water Supply Services Workers & 0\%   & 0\%   & 0\%   & 1\%   & 0\%   & 0\%   & 1\%   & 0\%   & 0\%   & 19\%  & 0\%   & 0\%   & 0\%   & 0\%   & 14\% \\
    \midrule
    Paper and Paper Products Production Processing Workers & 0\%   & 0\%   & 0\%   & 0\%   & 0\%   & 0\%   & 0\%   & 0\%   & 0\%   & 0\%   & 2\%   & 0\%   & 0\%   & 0\%   & 0\% \\
    \midrule
    Financial Servicess Workers & 0\%   & 0\%   & 0\%   & 0\%   & 1\%   & 63\%  & 0\%   & 0\%   & 0\%   & 0\%   & 0\%   & 0\%   & 0\%   & 0\%   & 0\% \\
    \midrule
    Specialized Equipment Manufacturing Workers & 0\%   & 0\%   & 0\%   & 0\%   & 0\%   & 0\%   & 0\%   & 0\%   & 0\%   & 0\%   & 1\%   & 0\%   & 0\%   & 0\%   & 0\% \\
    \midrule
    Instrumenta-tion Manufacturing Workers & 0\%   & 0\%   & 0\%   & 0\%   & 0\%   & 0\%   & 0\%   & 0\%   & 0\%   & 0\%   & 0\%   & 0\%   & 0\%   & 0\%   & 0\% \\
    \midrule
    Agricultural Technicians & 0\%   & 0\%   & 0\%   & 0\%   & 0\%   & 0\%   & 0\%   & 0\%   & 0\%   & 0\%   & 0\%   & 0\%   & 0\%   & 0\%   & 0\% \\
    \midrule
    Chemical Materials and Chemicals Manufacturing Workers & 0\%   & 0\%   & 0\%   & 0\%   & 2\%   & 0\%   & 0\%   & 0\%   & 0\%   & 1\%   & 1\%   & 0\%   & 0\%   & 0\%   & 0\% \\
    \midrule
    Pharmaceutical Manufacturing Workers & 0\%   & 0\%   & 0\%   & 0\%   & 0\%   & 0\%   & 0\%   & 0\%   & 0\%   & 0\%   & 0\%   & 1\%   & 0\%   & 0\%   & 0\% \\
    \midrule
    Electricity, Heat, Gas, Water Production and Distribution Workers & 0\%   & 0\%   & 0\%   & 0\%   & 0\%   & 0\%   & 0\%   & 1\%   & 0\%   & 30\%  & 0\%   & 0\%   & 0\%   & 0\%   & 5\% \\
    \midrule
    Electrical Machinery and Equipment Manufacturing Workers & 0\%   & 0\%   & 0\%   & 0\%   & 0\%   & 0\%   & 0\%   & 0\%   & 0\%   & 0\%   & 3\%   & 0\%   & 0\%   & 0\%   & 0\% \\
    \midrule
    Petroleum Processing and Coking, Coal Chemical Production Workers & 0\%   & 0\%   & 0\%   & 0\%   & 0\%   & 0\%   & 0\%   & 0\%   & 0\%   & 2\%   & 0\%   & 0\%   & 0\%   & 0\%   & 0\% \\
    \midrule
    Transporta-tion Equipment and General Engineering Machinery Operators and Related Workers & 0\%   & 0\%   & 0\%   & 0\%   & 0\%   & 0\%   & 3\%   & 0\%   & 2\%   & 0\%   & 1\%   & 0\%   & 0\%   & 5\%   & 0\% \\
    \midrule
    Mining Workers & 0\%   & 0\%   & 0\%   & 0\%   & 1\%   & 0\%   & 0\%   & 0\%   & 0\%   & 5\%   & 0\%   & 0\%   & 0\%   & 68\%  & 0\% \\
    \midrule
    Metal Smelting and Rolling Processing Workers & 0\%   & 0\%   & 0\%   & 0\%   & 0\%   & 0\%   & 0\%   & 0\%   & 0\%   & 1\%   & 3\%   & 0\%   & 0\%   & 0\%   & 5\% \\
    \midrule
    Railroad, Marine, and Aviation Equipment Manufacturing Workers & 0\%   & 0\%   & 0\%   & 0\%   & 0\%   & 0\%   & 0\%   & 0\%   & 0\%   & 0\%   & 1\%   & 0\%   & 0\%   & 0\%   & 0\% \\
    \midrule
    Aircraft and Marine Technicians & 0\%   & 0\%   & 0\%   & 0\%   & 0\%   & 0\%   & 0\%   & 0\%   & 0\%   & 0\%   & 0\%   & 0\%   & 0\%   & 0\%   & 0\% \\
    \midrule
    General Equipment Manufacturing Workers & 0\%   & 0\%   & 0\%   & 0\%   & 0\%   & 0\%   & 0\%   & 0\%   & 1\%   & 1\%   & 1\%   & 0\%   & 0\%   & 0\%   & 0\% \\
    \midrule
    Chemical Fiber Manufacturing Workers & 0\%   & 0\%   & 0\%   & 0\%   & 0\%   & 0\%   & 0\%   & 0\%   & 0\%   & 0\%   & 0\%   & 0\%   & 0\%   & 0\%   & 0\% \\
    \midrule
    Automotive Manufacturing Workers & 0\%   & 0\%   & 0\%   & 0\%   & 0\%   & 0\%   & 0\%   & 0\%   & 0\%   & 0\%   & 1\%   & 0\%   & 0\%   & 0\%   & 0\% \\
    \midrule
    Tobacco and Products Processing Workers & 0\%   & 0\%   & 0\%   & 0\%   & 0\%   & 0\%   & 0\%   & 0\%   & 0\%   & 0\%   & 0\%   & 0\%   & 0\%   & 0\%   & 0\% \\
    \midrule
    Fisheries Production Workers & 0\%   & 0\%   & 0\%   & 0\%   & 0\%   & 0\%   & 0\%   & 0\%   & 0\%   & 0\%   & 0\%   & 0\%   & 1\%   & 0\%   & 0\% \\
    \midrule
    Husbandry Production Workers & 0\%   & 0\%   & 0\%   & 0\%   & 0\%   & 0\%   & 0\%   & 0\%   & 0\%   & 0\%   & 0\%   & 0\%   & 1\%   & 0\%   & 0\% \\
    \bottomrule
\end{longtable}%
\end{center}
\begin{center}
\begin{longtable}{|p{15em}|c|}
    \caption{Proportion of Occupations with >50\% Consistent Scores for Different Models}
    \label{tab:proportion consistent scores} \\
    \toprule
    \textbf{Model} & \textbf{Proportion} \\
    \midrule
    GPT & 83.0\% \\
    \midrule
    GLM2 & 64.4\% \\
    \midrule
    InternLM & 83.2\% \\
    \bottomrule
\end{longtable}%
\end{center}

\renewcommand{\thesection}{Appendix E}
\section{Data Appendix}\label{appendix_e}

In this Appendix we provide further description on the corpus of online job vacancy postings collected by the City Data Group.
\subsection{Web Sources and Data Provider} 
Our corpus of online job vacancy postings is provided by the labour market and analytics company ‘The City Data Group’. The City Data Group has been scraping online job vacancy postings in China since January 2015.
Each job vacancy posting is scraped from the internet, including major online job market platforms in China such as zhaoping.com, 51job, 58.com, Ganji.com, Lagou.com, and Kanzhun. Lightcast actively audits their list of web sources to ensure data from new websites is on-boarded in a timely manner. One of the main competitive advantages of the City Data Group’s data product is the breadth of their sources. These data are often referred to in the literature as the ‘near universe’ of online job vacancy postings.
\subsection{Recruitment Data Processing Workflow} 
Once an online job vacancy posting is scraped, the City Data Group processes this data to produce the online posting dataset. A description the data processing workflow are:
When aggregating raw recruitment data into standardized formats, we processed different fields in distinct ways.
Fields Stored in Their Original or Nearly Original Forms
1.	Company Name;
2.	Industry; if it is stated.
3.	Benefits;
4.	Job Title: Stored as a string.
5.	Company Size (Lower and Upper Limits): measured in the number of employees. For instance, if a company size is listed as "10-99 employees," enscale\_lower is 10, and enscale\_upper is 99.
6.	Experience Requirements: measured as the minimum required years. 
7.	Recruitment Numbers.
8.	Job Descriptions: Stored as a string with full job descriptions, including: iob Responsibilities, job details, working conditions, etc.

\subsection{Errors Checking and Missing Information} 
Overall, the data product is a highly informative and accurate product, but we also acknowledge that any dataset with hundreds of millions of observations scraped from different sources will never be perfect. Both the structured data and the plain text data require a number of pre-processing steps and the use of algorithmic feature extraction, which in a very small number of cases produce errors (e.g. misclassification of occupations, truncation of plain text, presence of erroneous text). In the following we highlight some of the errors we have encountered, and discuss the strategies we employed to ensure our results remain robust to such issues.
Missing Values
A relevant value (e.g. the educational requirement for a job, the salary for a job) might be missing for at least two reasons: (i) the employer does not mention this explicitly in the text of the job ad, and (ii) the algorithm used to extract this feature from the text failed. By our double checking, missing values are almost entirely due to lack of information, and not poor feature extraction. 
Erroneous Plain Text
In a very small number of cases we observe that the plain text includes some parts of the website other than the job description.  We exclude these erroneous pain text.
Duplicate Recruitment Ads
In some cases, a recruitment advertisement will be released duplicated within the same platform or across platform. We define duplicate job adverts as job adverts from the same firm recruiting the same type of employee are repeated during the thirty-day window period. In our empirical analysis, esp. counting vacancy number, we consider this concern by adjusting duplicate recruitment ads. 
\subsection{Representativeness of Online Job Vacancy Postings} 
The City Data Group frequently reviews the representativeness of the job vacancy postings it scrapes, to ensure the information renders an accurate picture at least the online recruitment labor market. Obviously, the online recruitment labor market does not equal to the whole labor market; to gauge the nuance difference between online recruiting labor market and the overall labor market, the one precondition is a representative aggregate recruitment dataset. Since in China there is no representative recruitment dataset such as JOLTS in United States, therefore, we cannot address the issue of representative of online job recruitment.

\end{appendices}


\end{document}